\documentclass[twocolumn,twocolappendix,times,tighten]{aastex63}

\usepackage{graphicx}	% Including figure files
\usepackage{amsmath}	% Advanced maths commands
\usepackage{amssymb}	% Extra maths symbols
\usepackage{wasysym}    % Extra astronomy symbols
\usepackage{bm}         % Bold maths
\usepackage{xcolor}
\usepackage{etoolbox}
\usepackage{ulem}
\makeatletter % Workaround to only color the year for cite commands
  \patchcmd{\NAT@citex}
    {\@citea\NAT@hyper@{%
      \NAT@nmfmt{\NAT@nm}%
      \hyper@natlinkbreak{\NAT@aysep\NAT@spacechar}{\@citeb\@extra@b@citeb}%
      \NAT@date}}
    {\@citea\NAT@nmfmt{\NAT@nm}%
    \NAT@aysep\NAT@spacechar\NAT@hyper@{\NAT@date}}{}{}

  \patchcmd{\NAT@citex}
    {\@citea\NAT@hyper@{%
      \NAT@nmfmt{\NAT@nm}%
      \hyper@natlinkbreak{\NAT@spacechar\NAT@@open\if*#1*\else#1\NAT@spacechar\fi}%
        {\@citeb\@extra@b@citeb}%
      \NAT@date}}
    {\@citea\NAT@nmfmt{\NAT@nm}%
    \NAT@spacechar\NAT@@open\if*#1*\else#1\NAT@spacechar\fi\NAT@hyper@{\NAT@date}}
    {}{}
\makeatother

\newcommand\bmath[1]{\bm{#1}}
\newcommand\mathbfss[1]{\bm{\mathsf{#1}}}
\newcommand\Lsun{\text{L}_{\astrosun}} % requires the wasysym package
\newcommand\Arepo{\textsc{arepo}} % code name
\newcommand\ArepoRT{\textsc{arepo-rt}} % code name
\newcommand\ArepoMCRT{\textsc{arepo-mcrt}} % code name
\received{August 4, 2020}
\revised{October 21, 2020}
\accepted{October 22, 2020}
\submitjournal{ApJ}

\shorttitle{Monte Carlo Radiation Hydrodynamics in \Arepo}
\shortauthors{Smith et al.}
\graphicspath{{./}{figures/}}
\begin{document}

% \title{Template \aastex Article with Examples:
% v6.3\footnote{Released on June, 10th, 2019}}

\title{\ArepoMCRT: Monte Carlo Radiation Hydrodynamics on a Moving Mesh}

\correspondingauthor{Aaron Smith; \href{mailto:arsmith@mit.edu}{arsmith@mit.edu}}
% \email{arsmith@mit.edu}

\author[0000-0002-2838-9033]{Aaron Smith}
\altaffiliation{NHFP Einstein Fellow.}
\affiliation{Department of Physics,
  % Kavli Institute for Astrophysics and Space Research,
  Massachusetts Institute of Technology, Cambridge, MA 02139, USA}

\author[0000-0001-6092-2187]{Rahul Kannan}
\affiliation{Center for Astrophysics, Harvard $\&$ Smithsonian, 60 Garden Street, Cambridge, MA 02138, USA}

\author[0000-0002-6543-2993]{Benny T.-H. Tsang}
\affiliation{Kavli Institute for Theoretical Physics, University of California, Santa Barbara, CA 93106, USA}

\author[0000-0001-8593-7692]{Mark Vogelsberger}
\affiliation{Department of Physics,
  % Kavli Institute for Astrophysics and Space Research,
  Massachusetts Institute of Technology, Cambridge, MA 02139, USA}

\author[0000-0003-3308-2420]{R\"{u}diger Pakmor}
\affiliation{Max-Planck Institute for Astrophysics, Karl-Schwarzschild-Str.~1, D-85741 Garching, Germany}

% \collaboration{1}{(AAS Journals Data Scientists collaboration)}

% \author{Butler Burton}
% \affiliation{Leiden University}
% \affiliation{AAS Journals Associate Editor-in-Chief}
% \nocollaboration{1}

% \author{Amy Hendrickson}
% \altaffiliation{AASTeX v6+ programmer}
% \affiliation{TeXnology Inc.}

% \collaboration{1}{(LaTeX collaboration)}

% \author{Julie Steffen}
% \affiliation{AAS Director of Publishing}
% \affiliation{American Astronomical Society \\
% 1667 K Street NW, Suite 800 \\
% Washington, DC 20006, USA}

% \author{Scott Chernoff}
% \affiliation{IOP Publishing, Washington, DC 20005}

% \nocollaboration{2}

%% Note that the \and command from previous versions of AASTeX is now
%% depreciated in this version as it is no longer necessary. AASTeX
%% automatically takes care of all commas and "and"s between authors names.

%% AASTeX 6.3 has the new \collaboration and \nocollaboration commands to
%% provide the collaboration status of a group of authors. These commands
%% can be used either before or after the list of corresponding authors. The
%% argument for \collaboration is the collaboration identifier. Authors are
%% encouraged to surround collaboration identifiers with ()s. The
%% \nocollaboration command takes no argument and exists to indicate that
%% the nearby authors are not part of surrounding collaborations.

%% Mark off the abstract in the ``abstract'' environment.
\begin{abstract}

We present \ArepoMCRT, a novel Monte Carlo radiative transfer (MCRT) radiation-hydrodynamics (RHD) solver for the unstructured moving-mesh code \Arepo. Our method is designed for general multiple scattering problems in both optically thin and thick conditions. We incorporate numerous efficiency improvements and noise reduction schemes to help overcome efficiency barriers that typically inhibit convergence. These include continuous absorption and energy deposition, photon weighting and luminosity boosting, local packet merging and splitting, path-based statistical estimators, conservative (face-centered) momentum coupling, adaptive convergence between time steps, implicit Monte Carlo algorithms for thermal emission, and discrete-diffusion Monte Carlo techniques for unresolved scattering, including a novel advection scheme. We primarily focus on the unique aspects of our implementation and discussions of the advantages and drawbacks of our methods in various astrophysical contexts. Finally, we consider several test applications including the levitation of an optically thick layer of gas by trapped infrared radiation. We find that the initial acceleration phase and revitalized second wind are connected via self-regulation of the RHD coupling, such that the RHD method accuracy and simulation resolution each leave important imprints on the long-term behavior of the gas.

\end{abstract}

%% Keywords should appear after the \end{abstract} command.
%% See the online documentation for the full list of available subject
%% keywords and the rules for their use.
% \keywords{editorials, notices ---
% miscellaneous --- catalogs --- surveys}
\keywords{radiative transfer --- radiation: dynamics --- methods:numerical}
% -- galaxies: formation -- galaxies: high-redshift

%% From the front matter, we move on to the body of the paper.
%% Sections are demarcated by \section and \subsection, respectively.
%% Observe the use of the LaTeX \label
%% command after the \subsection to give a symbolic KEY to the
%% subsection for cross-referencing in a \ref command.
%% You can use LaTeX's \ref and \label commands to keep track of
%% cross-references to sections, equations, tables, and figures.
%% That way, if you change the order of any elements, LaTeX will
%% automatically renumber them.
%%
%% We recommend that authors also use the natbib \citep
%% and \citet commands to identify citations.  The citations are
%% tied to the reference list via symbolic KEYs. The KEY corresponds
%% to the KEY in the \bibitem in the reference list below.

\section{Introduction}
\label{sec:intro}
Observations of light at all wavelengths of the electromagnetic spectrum have been essential to our understanding of the cosmos and the wide range of astrophysical phenomena found therein. Intense radiation from stars and black holes can also affect the dynamical evolution of small- to large-scale systems as sources of thermal, mechanical, and chemical feedback. A substantial effort has been made to model the interplay between gas and radiation with robust theoretical models and numerical techniques. In many cases, the multiscale, multiphysics nature of such problems necessitates fully coupled radiation-hydrodynamics (RHD) simulations to precisely capture the underlying physics \citep{Pomraning1973,Mihalas1984,Castor2004}.

The large variety of environments requiring the study of RHD and decoupled radiative transfer has led to the development of different algorithms with advantages for specialized applications. Part of the complexity is the high dimensionality of the radiation field, which can vary as a function of space, direction, frequency, and time. Most schemes employed by modern codes can broadly be classified as: (i) directly integrating the transport equation along characteristic paths; (ii) solving discrete-ordinate representations of the transport equation; and (iii) solving reduced-dimensionality angular-moment equations derived with an approximate closure relation. Each of these, or hybrid combinations thereof, have been highly successful in exploratory and precision studies in astrophysics. A few contemporary methods for multidimensional RHD include flux-limited diffusion \citep[FLD;][]{Levermore1981,Turner2001,Krumholz2007,Commercon2011}, first moment closure \citep[M1;][]{Levermore1984,Gonzalez2007,Rosdahl2013,Skinner2013,Kannan2019}, variable Eddington tensor formulations \citep[VET;][]{Stone1992,Davis2012,Jiang2012,Jiang2014}, optically thin VET \citep[OTVET;][]{Gnedin2001}, adaptive ray-tracing \citep{Abel1999,Whalen2006,Trac2007,Wise2011,Rosen2017}, transport in light cones \citep{Pawlik2008}, and the method of characteristics moment closure \citep[MOCMC;][]{Ryan2020}. The community continues to benefit from numerous contributions enabling further radiative transfer studies.

In this paper, we present a Monte Carlo radiative transfer (MCRT) implementation for the moving-mesh hydrodynamics code \Arepo\ \citep{Springel2010,Pakmor2011,Pakmor2016,Weinberger2020}. Our efforts are complementary to other RHD methods already implemented in the code, including direct discretization \citep{Petkova2011}, SIMPLEX triangulation \citep{Jaura2018}, and M1 closure \citep{Kannan2019}. The MCRT method provides an accurate approach to modeling radiation fields by sampling from physically motivated probability distribution functions \citep[for a recent review see][]{Noebauer2019}. Complex phenomena arising from microphysical processes, such as multiple scattering and frequency redistribution, can be accounted for from first principles and astrophysical observables can be assembled one photon packet at a time. This often leads to conceptually simple implementations and interpretations of the method. One of the main drawbacks of MCRT is the noise from photon packet discretization. However, with enough computational resources and efficiency-improving algorithms, convergence can always be obtained to arbitrarily high accuracy. Quantities derived from binned photon statistics typically suffer from a slow convergence rate with the number of samples as $1/\sqrt{N}$, although this is independent of the dimensionality of the physical setup.

Indeed, MCRT constitutes a valuable alternative approach to RHD that can outperform other methods in terms of accuracy and emergent statistics in many computationally demanding situations. It is also clear that the MCRT RHD approach is maturing with an increasing number of codes and applications \citep{Nayakshin2009,Noebauer2012,Cleveland2015,Harries2015,Noebauer2015,Roth2015,Ryan2015,Tsang2015,Harries2017,SmithRHD2017,Tsang2018,Vandenbroucke2018}. Therefore, it is timely to incorporate on-the-fly and post-processing MCRT into \Arepo\ for accurate native modeling of radiation fields and emergent observables. Especially given the demonstrated capabilities of \Arepo\ for studying astrophysical problems over the past decade \citep[for a brief but comprehensive list of applications see][]{Weinberger2020}. Additional access to accurate radiative transfer on a moving mesh can help facilitate important insights in astronomy spanning a wide range of scales, large and small.

Although many of the algorithms discussed in this work are new, we have adopted or adapted most of the concepts from the mature MCRT literature. For example, implicit Monte Carlo \citep[IMC;][]{FleckCummings1971,Brooks1986,Brooks1989} algorithms provide an essential framework for stable and efficient time-dependent coupling of the radiative transfer and thermal energy equations by treating a portion of absorption followed by re-emission as effective scattering. However, there are tradeoffs with implicit schemes; for example, spurious behavior has been found in extreme cases which can only be mitigated with corrective IMC or iterative variants \citep{Long2014,Cleveland2018}. More importantly, in astrophysical applications MCRT photon packets become diffusive and inefficient to track without accelerated transport. One option is to maintain continuous photon positions with movement in optically thick zones according to a modified random walk \citep{FleckCanfield1984}. This technique is widely used and has undergone several recent improvements \citep{Min2009,Robitaille2010,Keady2017}.

In practice, random walks are not always optimal because they become inactive near cell interfaces. Still, such diffusion approximations are necessary to obtain accurate MCRT results at high optical depths \citep{Camps2018}. Therefore, discrete-diffusion Monte Carlo (DDMC) techniques have been developed to increase the efficiency of MCRT calculations in opaque media \citep{Gentile2001,Densmore2007}. The idea is to replace many unresolved scatterings with a single jump to a neighboring cell based on a discretized diffusion equation. Under the Fick's law closure relation of the radiative transfer equation (RTE), the diffusive term operates as spatial leakage across cell interfaces that is naturally incorporated into the Monte Carlo (MC) paradigm. A robust hybrid scheme allows the conversion between DDMC and MC particles for accurate propagation through optically thin cells as well. The DDMC method has also been extended to incorporate frequency-dependent transfer \citep{Abdikamalov2012,Densmore2012,Wollaeger2013,Wollaeger2014}.

The paper is organized as follows. In Section~\ref{sec:methods}, we discuss the main numerical methodology, emphasizing aspects that are unique to our implementation, hereafter known as \ArepoMCRT. In Section~\ref{sec:tests}, we test our implementation against known analytic and numerical solutions from the literature. In Section~\ref{sec:levitation}, we apply the code to the problem of radiative forcing of a dusty atmosphere. Finally, in Section~\ref{sec:summary}, we provide a summary of findings and discussion of future applications.

\section{Numerical Methodology}
\label{sec:methods}
In this section, we describe our Monte Carlo radiative transfer RHD implementation, which is fully integrated into the moving-mesh code \Arepo\ \citep{Springel2010}. Specifically, \Arepo\ employs a second-order finite-volume method to solve the ideal magneto-hydrodynamical equations on an unstructured Voronoi tessellation, which is free to move with the local fluid velocity. Native ray-tracing through three-dimensional Voronoi tessellations is well understood in computational geometry and has been employed by several post-processing MCRT codes with great success \citep{Camps2013,CampsBaes2015,Hubber2016,SmithDCBH2017}. We therefore focus on physics rather than geometry in this section.

\subsection{Radiation transport}
\label{sec:RT}
The specific intensity $I_\nu(\bmath{r}, \bmath{n}, t)$ encodes all information about the radiation field taking into account the frequency $\nu$, spatial position $\bmath{r}$, propagation direction unit vector $\bmath{n}$, and time $t$. The general RTE is given in the lab frame by
\begin{equation} \label{eq:RTE}
  \frac{1}{c} \frac{\partial I_\nu}{\partial t} + \bmath{n} \bmath{\cdot} \bmath{\nabla} I_\nu = j_\nu - k_\nu I_\nu \, ,
\end{equation}
where $k_\nu$ is the absorption coefficient and $j_\nu$ is the emission coefficient. We define the mean intensity as $J_\nu \equiv \frac{1}{4\pi} \int \text{d}\Omega I_\nu$, which is related to the energy density by $u_\nu = \frac{4\pi}{c} J_\nu$. To avoid confusion, in this paper we denote the gas internal energy density by $u_\text{g}$.

In this paper we restrict the discussion to monochromatic radiation, although this is not required in the code and will be relaxed in future applications. In this context a convenient parameterization for the absorption coefficient is through a constant opacity $\kappa \equiv k / \rho$ and scattering albedo $A \equiv k_\text{s} / (k_\text{s} + k_\text{a})$ where $\rho$ denotes the gas density and $k_\text{s}$ and $k_\text{a}$ are the purely scattering and absorbing components. Furthermore, if we assume local thermodynamic equilibrium (LTE) then the right-hand side of equation~(\ref{eq:RTE}) simplifies to include (i) dust absorption and thermal emission as $k_\text{a} (B - I)$ based on the Planck function $B$ at temperature $T$, (ii) isotropic elastic scattering as $\frac{k_\text{s}}{4\pi} \int [I(\bmath{n}') - I(\bmath{n})] \text{d}\bmath{n}' = k_\text{s} (J - I)$, and (iii) external sources $j_\text{ext}$, e.g. geometrically assigned sources. For completeness, the Planck (blackbody) distribution is defined as $B_\nu(T) \equiv \frac{2 h \nu^3}{c^2} (e^{h\nu/k_\text{B}T} - 1)^{-1}$ with a normalization such that the energy density that radiation would have if it were in thermodynamic equilibrium with gas is $u_\text{r} = \frac{4\pi}{c} \int B_\nu(T)\,\text{d}\nu = a_\text{B} T^4$. Finally, the Planck mean absorption coefficient is $k_\text{P} = \int k_{\nu,\text{a}} B_\nu(T)\,\text{d}\nu / \int B_\nu(T)\,\text{d}\nu$, which is trivially $k_\text{a}$ under the assumption of a frequency-independent (gray) opacity. Thus, we summarize the simplified version of equation~(\ref{eq:RTE}) as
\begin{equation} \label{eq:simplified-RTE}
  \frac{1}{c} \frac{\partial I}{\partial t} + \bmath{n} \bmath{\cdot} \bmath{\nabla} I = k_\text{a} (B - I) + k_\text{s} (J - I) + j_\text{ext} \, .
\end{equation}
The radiation flux and pressure are given by $\bmath{F}_\nu = \int \text{d}\Omega I_\nu \bmath{n}$ and $\mathbfss{P}_\nu \equiv c^{-1} \int \text{d}\Omega I_\nu \bmath{n} \otimes \bmath{n}$, respectively. Therefore, the moment equations provide a general framework for radiation--gas coupling:
\begin{equation} \label{eq:simplified-RTE-0}
  \frac{\partial u}{\partial t} + \bmath{\nabla} \bmath{\cdot} \bmath{F} = c k_\text{a} (u_\text{r} - u)
\end{equation}
and
\begin{equation} \label{eq:simplified-RTE-1}
  \frac{1}{c^2} \frac{\partial \bmath{F}}{\partial t} + \bmath{\nabla} \bmath{\cdot} \mathbfss{P} = -\frac{k \bmath{F}}{c} \, .
\end{equation}
For presentation purposes we do not show moments of $j_\text{ext}$.

\subsection{Radiation hydrodynamics}
\label{sec:RHD}
The equations governing nonrelativistic hydrodynamics can be written in an Eulerian reference frame as a set of conservation laws for mass, momentum, and total energy \citep{Castor2004}:
\begin{equation}
  \frac{\partial \rho}{\partial t} + \bmath{\nabla} \bmath{\cdot} (\rho \bmath{v}) = 0 \, ,
\end{equation}
\begin{equation}
  \frac{\partial \rho \bmath{v}}{\partial t} + \bmath{\nabla} \bmath{\cdot} (\rho \bmath{v} \otimes \bmath{v}) + \bmath{\nabla}P = \frac{k \bmath{F}}{c} \, ,
\end{equation}
and
\begin{equation}
  \frac{\partial \rho e}{\partial t} + \bmath{\nabla} \bmath{\cdot} \left[(\rho e + P) \bmath{v}\right] = c k_\text{a} (u - u_\text{r}) + \bmath{v} \bmath{\cdot} \frac{k \bmath{F}}{c} \, .
\end{equation}
Here, $\rho$ is the density, $\bmath{v}$ the velocity, $P$ the pressure, $e \equiv \epsilon + \frac{1}{2} |\bmath{v}|^2$ the total specific energy. We typically assume an ideal gas equation of state so the pressure is isotropic and specified by $P = (\gamma_\text{ad} - 1) u_\text{g}$, where $u_\text{g} = \rho \epsilon$ and $\gamma_\text{ad} \equiv C_P / C_V$ is the adiabatic index, or ratio of specific heat at constant pressure to that at constant volume. These equations would also be modified in the presence of additional physics, such as magnetic fields or self-gravity, which is already present in the \Arepo\ code \citep{Springel2010,Weinberger2020}.

In the current work we primarily focus on the MCRT and RHD physics, noting that we expect to improve the implementation as needed for future applications. For now, the radiation is coupled to the hydrodynamics with standard operator splitting that is formally first order in time and space. While second-order schemes for spatial emission sampling and path integration utilizing the gradient information in \Arepo\ are possible, we leave the exploration of potential benefits to future work. At this point we also require global time-stepping, but plan to allow compatibility with the individual time-stepping scheme of \Arepo, i.e. the efficient factor of two hierarchy of timesteps, which would then also benefit from a second-order time-integration scheme.

The energy and momentum exchanged between gas and radiation is performed at the end of each time step as follows. Firstly, emission processes can remove internal energy from the gas, which is converted to radiation as MC photon packets. Secondly, the MCRT transport calculations are performed, meanwhile tracking the cumulative interactions with gas due to photon absorption and scattering processes. Lastly, the conserved gas quantities are collectively updated to reflect the net exchange. In the next section we describe the IMC technique to treat the source and transport steps (semi-)implicitly. The mesh configuration remains unchanged during transport, however photon packets generally persist across time steps. This is achieved by saving the particle positions along with the host cell indices, which are validated before transport via a breadth-first face neighbor walk. The photon packets are also efficiently exchanged between tasks during a domain decomposition. Nonlocal transport is handled by collecting photons at domain boundaries and then sending batches via asynchronous point-to-point communication patterns \citep[similar to][]{Rosen2017}. Additional details of the implementation are described in the remaining subsections.

\subsection{Implicit Monte Carlo}
\label{sec:IMC}
Under LTE conditions, radiation and gas are tightly coupled through the strong temperature dependence of the thermal emission. The IMC algorithm employs a semi-implicit time discretization to transform the nonlinear thermal RTE into a system of linearized equations naturally incorporated into the MC method. Effectively, the method replaces a portion of absorption and re-emission with elastic scattering, thus reducing the amount of quasi-equilibrium energy exchange between gas and radiation. We refer the interested reader to \citet{FleckCummings1971} and \citet{Wollaber2016} for detailed derivations and numerical discussions. For our purposes, we introduce the material equation as (simplified following equation~\ref{eq:simplified-RTE})
\begin{equation} \label{eq:material-equation}
  \frac{1}{c} \frac{\partial u_\text{g}}{\partial t} = \frac{1}{c\beta} \frac{\partial u_\text{r}}{\partial t} = k_\text{a} (u - u_\text{r}) \, ,
\end{equation}
where $\beta \equiv \partial u_\text{r}/\partial u_\text{g}$ quantifies the thermal coupling. We then expand $u_\text{r}$ to first order in time for $t \in [t^n,t^{n+1}]$
\begin{equation} \label{eq:ur-first-order}
  u_\text{r}(t) \simeq u_\text{r}^n + \alpha \Delta t (\partial u_\text{r} / \partial t)^n \, .
\end{equation}
The parameter $\alpha \in [0,1]$ is a coefficient that interpolates between a fully explicit ($\alpha = 0$) and implicit ($\alpha = 1$) scheme for updating $u_\text{r}$ over the time step $\Delta t$. Substituting $u_\text{r}(t)$ from equation~(\ref{eq:ur-first-order}) into (\ref{eq:material-equation}) and eliminating $(\partial u_\text{r} / \partial t)^n$ gives the following:
\begin{equation} \label{eq:ur-first-order-fleck}
  u_\text{r}(t) = f u_\text{r}^n + (1 - f) u \, ,
\end{equation}
in which we have introduced the so-called Fleck factor
\begin{equation} \label{eq:fleck}
  f \equiv \frac{1}{1 + \alpha c \Delta t \beta k_\text{P}} = \frac{1}{1 + 4 \alpha c\Delta t (1-A) \kappa u_\text{r}/\epsilon} \, ,
\end{equation}
where in the second equality we assume a constant scattering albedo $A$, opacity $\kappa$, and an ideal gas equation of state such that $\beta = 4 u_\text{r} / u_\text{g}$. Finally, we substitute $u_\text{r}$ from equation~(\ref{eq:ur-first-order-fleck}) into equation~(\ref{eq:simplified-RTE}) to obtain the implicit radiation transport equation:
\begin{equation} \label{eq:simplified-RTE-fleck}
  \frac{1}{c} \frac{\partial I}{\partial t} + \bmath{n} \bmath{\cdot} \bmath{\nabla} I = k_\text{ea} (B - I) + k_\text{es} (J - I) + j_\text{ext} \, ,
\end{equation}
where the effective absorption and scattering coefficients are
\begin{equation} \label{eq:k_ea}
  k_\text{ea} \equiv f k_\text{a}
\end{equation}
and
\begin{equation}
  k_\text{es} \equiv k_\text{s} + (1 - f) k_\text{a} \, .
\end{equation}
Thus, after calculating the Fleck factor the numerical radiation transport coefficients are modified to reflect the replacement of a portion of thermal absorption with scattering. For notational simplicity, we drop the `effective' subscript with the understanding that the IMC scheme is implied throughout.

\subsection{Monte Carlo procedure}
\label{sec:MC}
Under the MCRT paradigm we represent the specific intensity with an ensemble of photon packets, which are each characterized by an energy weight $\varepsilon_k$, position $\bmath{r}_k$, normalized direction $\bmath{n}_k$, and time $t_k$, where the index $k$ refers to an individual MC packet. We emphasize that in our implementation different MC photon packets can each have varying energies, which can greatly improve the sampling and convergence statistics. \Arepo\ employs a finite-volume method to solve the gas conservation equations so in the first-order scheme we assume constant gas properties within each cell. We treat the thermal emission term $k_\text{a} B$ in equation~(\ref{eq:simplified-RTE}) as a continuous process for the gas but stochastically for radiation. The total thermal radiation energy emitted by each cell assuming gray opacity is
\begin{equation}
  \Delta E_{\text{em},i} = c \Delta t V_i k_\text{a} a_\text{B} T^4 \, ,
\end{equation}
where $\Delta t$ denotes the time step and $V_i$ is the current volume of cell $i$.
We note that with IMC the emission term is corrected by the Fleck factor as inherited from equation~(\ref{eq:k_ea}). For numerical stability, we subtract this exact energy from each gas cell, however the insertion of MC packets is necessarily discretized. This is achieved by constructing the cumulative distribution function from individual cells, drawing a random number to find the emitting cell, and assigning the photon packet position uniformly within the Voronoi cell volume. The emission direction is isotropic in the comoving frame of the gas and the time is uniformly distributed over the duration of the time step, $t_k \in [t_n,t_{n+1}]$. Furthermore, we incorporate an emission weighting scheme to accelerate the convergence of MC sampling. This allows simulations to dynamically allocate photon packets to optimally sample the radiation field, which we currently implement as a power law luminosity boost $\propto L^\gamma$ properly normalized to conserve energy. Specifically, $\gamma \in (0,1)$ alleviates the $1 / N_\text{ph}$ sensitivity limit inherent to uniform sampling, thereby reducing emissivity disparities that often lead to statistical noise in dim regions and unnecessary oversampling in bright regions. If the total radiation energy at a given time is $\mathcal{E}_n$ then the photon weight is $w_k \equiv \varepsilon_k / \mathcal{E}_n$, so by construction $\sum w_k = 1$.

The subsequent transport of photon packets follows the usual MC procedure for scattering and escape \citep[e.g.][]{Smith2015,Tsang2015}. We determine the optical depth to scattering based on an exponential distribution, i.e. $\tau_\text{scat} = -\ln \zeta$ where $\zeta$ is a random number uniformly distributed in $[0,1]$. The location of the scattering interaction is calculated implicitly via piecewise constant integration, i.e. by continually moving the photon through each cell until $\tau_\text{scat}$ is exhausted. After each scattering event the photon is assigned a new direction $\bmath{n}_k$ and the ray-tracing procedure continues until the photon (i) reaches the end of the time step, (ii) arrives at a specified escape criterion, or (iii) is removed by an absorption process (see Section~\ref{sec:MC_Estimators}). Each time the photon moves a distance $\Delta \ell_k$ there is a corresponding change in position of $\Delta \bmath{r}_k = \Delta \ell_k \bmath{n}_k$, elapsed time of $\Delta t_k = \Delta \ell_k / c$, and traversed optical depth of $\Delta \tau_k = k_\text{s} \Delta \ell_k$.

\subsection{Photon splitting and merging}
The instantaneous representation of the MCRT radiation field is the result of nontrivial sourcing and transport. In addition, photons persist across time steps, which can lead to dense accumulations of packets in regions of $(\bmath{r}, \bmath{n}, \nu)$ phase space. We therefore implement photon splitting and merging. Specifically, at the beginning of each time step we split statistically overweight packets, e.g. down to a few standard deviations above the mean, $\varepsilon_k \gtrsim \langle \varepsilon \rangle + f_\text{split} \langle \langle \varepsilon^2 \rangle - \langle \varepsilon \rangle^2 \rangle^{1/2}$. This simple prescription efficiently regulates the distribution of packet weights during transport calculations to reduce unnecessary variance. We note that in scattering media redundant photons quickly disperse to better sample alternative trajectories. In the future, particular applications may require additional splitting criteria to increase the signal-to-noise in optically thin regions or at large distances from the last scattering or emission event \citep[e.g. similar to][]{Harries2015}.

We also implement optional photon merging schemes to combine underweight photons, remove redundant information, or reduce the total number of photons in the simulation. As packets are already assigned to gas cells, at the end of each time step we perform a linear sort to count and identify merger candidates as groups of packets sharing a common host cell. This has the advantage of being computationally efficient and ensuring that the merged packets remain within the convex hull of the Voronoi cells. We currently consider all photons in the same cell as merger candidates, but one might include additional criteria such as the photon weight in the merge condition.

Once we have a list of photons in each cell we can further bin them into directional groups to preserve angular information after grouping. However, binning can lead to numerical artifacts by biasing paths toward certain directions. Therefore, our preferred implementation iteratively merges the pair of photons with closest angular separation, until all pairs exceed a threshold angle, e.g. $\theta > 30\degr$. This has the advantage of being unique and not favoring any particular directions. Although our algorithm is technically $\mathcal{O}(N^2)$, a matrix-like data structure means one particle's row and column is removed while the other's is reused for the merged packet. That is, we only recompute dot products for the row and column of the new packet, retaining angles for unaffected pairs and shrinking the effective matrix dimensions. In practice the merging calculations are a negligible fraction of the overall simulation runtime. However, when this is not the case, we optionally introduce a short-circuiting condition during the pair search to immediately merge sufficiently close photons, e.g. $\theta < 5\degr$, and limit the remaining candidates.

While merging can result in information loss, it is straightforward to characterize. Specifically, the merger only preserves the group momentum and center of energy position\footnote{Alternatively, one might prioritize conserving the total energy, which results in a fractional gain of momentum compared to the original radiation field. Beyond these simple choices, one might leverage the insight that particle data is often redundant to obtain a weighted subset (called a \textit{coreset}) to replace direct merging with a consolidated representation of the intensity.}. For a pair of photons the new energy is $\varepsilon' = (\varepsilon_1^2 + \varepsilon_2^2 + 2 \varepsilon_1 \varepsilon_2 \cos\theta)^{1/2}$, which for equal weight photons results in a fractional loss of $f_\text{loss} = 1 - \varepsilon' / (\varepsilon_1 + \varepsilon_2) = 1 - \frac{1}{2} \sqrt{2 + 2 \cos\theta} \approx \frac{1}{8} \theta^2 + \mathcal{O}(\theta^4)$, such that the maximum is $f_\text{loss} \lesssim 3.4\%\,(\theta/30\degr)^2$. The final outcome of multiple particle mergers is more complex but can be bounded by considering a ring configuration, i.e. equal weights and maximal separations, resulting in $f_\text{loss} \leq 1 - \cos\theta \approx \frac{1}{2} \theta^2 = 13.7\%\,(\theta/30\degr)^2$. More realistic is the case of many particles isotropically distributed over a spherical cap, which can be derived via a surface rotation as half of the loss in the ring case, or $f_\text{loss} \approx 6.9\%\,(\theta/30\degr)^2$. Unfortunately, these losses are unavoidable when employing splitting and merging to help regulate the weight distribution and avoid skewed or imbalanced statistics.

\subsection{Continuous absorption}
\label{sec:MC_Estimators}
We employ the continuous absorption method to minimize MC sampling noise. Treating the $-k_\text{a} I$ term in equation~(\ref{eq:RTE}) deterministically is a variance-reduction technique that accelerates the convergence of the radiation field by adjusting the weight of photon packets. Thus, after considering all other event-based processes, such as scattering or grid traversal, the photon energy weight is reduced by $e^{-\tau_\text{a}}$, where for notational simplicity we use $\tau_\text{a} \equiv k_\text{a} \Delta \ell_k$. Photon packets with negligible weight can be eliminated by applying a threshold condition, e.g. $w_k > w_\text{min} \sim 10^{-12}$. Furthermore, the internal energy deposited to the gas is
\begin{equation} \label{eq:energy_deposition}
  \Delta E_{\text{abs},i} = \sum_\text{paths} \varepsilon_k (1 - e^{-\tau_\text{a}}) \, ,
\end{equation}
where the sum is over all photon paths within cell $i$. We also employ a path-based estimator for the radiation energy density specifically accounting for the decreasing energy contribution due to continuous absorption \citep{SmithTsang2018}. Otherwise the energy and momentum deposition can be overestimated, especially in cases where the absorption optical depth is greater than unity. The correction factor is $\int_0^{\tau_\text{a}} e^{-\tau'_\text{a}}\,\text{d}\tau'_\text{a}/\tau_\text{a} = (1 - e^{-\tau_\text{a}}) / \tau_\text{a}$, or approximately $1 - \tau_\text{a}/2$ in the optically thin limit for absorption, which is important for numerical stability although we use the exact expression in the equations that follow. The corresponding corrected energy density based on the total residence time of propagating photons is \citep{Lucy1999}
\begin{equation} \label{eq:residence_energy}
  u_i = \sum_\text{paths} \frac{\varepsilon_k}{V_i} \frac{(1 - e^{-\tau_\text{a}})}{k_\text{a} c\Delta t} \, ,
\end{equation}
with the sum again over all paths within cell $i$.

\subsection{Momentum coupling}
\label{sec:momentum}
The momentum deposition is also path-based and includes both absorption and scattering contributions as
\begin{equation} \label{eq:momentum}
  \Delta \bmath{p}_i = \sum_\text{paths} \frac{\tau \varepsilon_k}{c} \left(\frac{1 - e^{-\tau_\text{a}}}{\tau_\text{a}}\right) \bmath{n}_k \, .
\end{equation}
Furthermore, as scattering occurs in the comoving frame the following kinetic energy is also transferred to the gas:
\begin{equation} \label{eq:kinetic_energy}
  \Delta E_{\text{kin},i} = \bmath{v} \bmath{\cdot} \Delta \bmath{p}_i \, .
\end{equation}
We note that momentum coupling in non-Cartesian coordinates is complicated by the fact that the reference unit vectors can change along traversed paths. In Appendix~\ref{sec:spherical_coordinates}, we provide additional discussion specific to calculations in spherical geometry, which is commonly used in astrophysical simulations.

We have implemented two methods for momentum coupling that incorporate different meanings to the summations in equation~(\ref{eq:momentum}). The first assigns the momentum directly to the host cell, i.e. the usual volume-integrated coupling. However, this can underestimate the radiation pressure force in the immediate vicinity around unresolved sources and therefore the impact on gas properties \citep{Hopkins2019}. It is often impractical to resolve photon mean free paths in hydrodynamics simulations so we provide a general Monte Carlo solution to conserve momentum in such cases. Our approach is a neighbor-based method, which also applies momentum to the cell that the photon packet would enter next. In the case of multiple scattering, momenta from each segment of the overall trajectory are imparted independently onto different neighbors according to individual path-based estimators. The schemes we propose are also particularly well suited for the unstructured meshes encountered in \Arepo, and adhere to the MCRT philosophy of accurately capturing sub-grid physical interactions\footnote{We recognize that alternative approaches could give qualitatively similar results. For example, the RTE might admit an exact or approximate solution including only local sources, which would provide the initial conditions for nonlocal MCRT with standard volume-integrated momentum coupling. On the other hand, one might treat self-canceling momentum as a source of turbulent energy or pressure in the hydrodynamics equations.}.

We emphasize that both volume- and face-integrated methods can lead to unphysical results for isotropic sources embedded within optically thick cells. For example, a constant luminosity point source with negligible absorption produces an inverse square law flux $\bmath{F} = L \hat{\bmath{r}} / (4 \pi r^2)$, such that the energy density for a pure scattering medium is given by $\bmath{F} = -c \bmath{\nabla} u / 3k$ and $u = 3 k L/ (4 \pi c r)$. By employing Gauss's theorem the integrated momentum rate is
\begin{equation} \label{eq:momentum_rate}
  \dot{\bmath{p}}_\text{tot} = \int \frac{k \bmath{F}}{c} \text{d}V = \oint \frac{u}{3} \text{d}\bmath{A} = \oint \frac{k L}{4 \pi c r} \text{d}\bmath{A} \, .
\end{equation}
If the source is located near a cell center then the net momentum is zero even though the correct radially outward value, corresponding to a sphere with differential surface area $\text{d}\bmath{A} = r^2 \text{d}\Omega \hat{\bmath{r}}$, is
\begin{equation}
  \dot{p}_{r,\text{tot}} = \left[\dot{\bmath{p}}_\text{tot} \bmath{\cdot} \hat{\bmath{r}} \right]_{r = \text{const}} = \frac{\tau(<r) L}{c} \, .
\end{equation}
The neighbor method conserves the total momentum because the integration is always directed toward an interface with positive vector orientation. However, as discussed by \citet{Hopkins2019}, by representing cell interfaces as planes with finite area a fraction of momentum is lost due to residual vector cancellation in the transverse directions. To gain intuition, we approximate cells as having $N \gg 1$ faces subtending equal solid angles of $4 \pi / N$. We further employ the small-angle approximation and approximate each face as a disk at a distance $z$ with relative differential surface area $\text{d}\bmath{A} / z^2 = 2 \pi \varphi \text{d}\varphi \hat{\bmath{z}}$ and maximum conic opening angle $\varphi_\text{max} \approx 2 / \sqrt{N}$. Thus, the total relative momentum rate is
\begin{equation}
  \frac{\dot{p}_{N,\text{tot}}}{\dot{p}_{r,\text{tot}}} \approx \frac{N}{2} \int_0^{\varphi_\text{max}} \frac{\varphi\,\text{d}\varphi}{\sqrt{1 + \varphi^2}} \approx 1 - \frac{1}{N} + \mathcal{O}\left(N^{-2}\right) \, ,
\end{equation}
so a Voronoi tessellation with a larger number of neighbors will have a smaller geometric loss than a regular Cartesian grid. We note that even when the point source is close to a cell boundary the geometric efficiency is still greater than $\dot{p}_{z,\text{tot}}/\dot{p}_{r,\text{tot}} \geq \int_0^1 \rho \sqrt{1-\rho^2}\,\text{d}\rho / \int_0^1 \rho\,\text{d}\rho = 2/3$ there.

\begin{figure}
\centering
\includegraphics[width=\columnwidth]{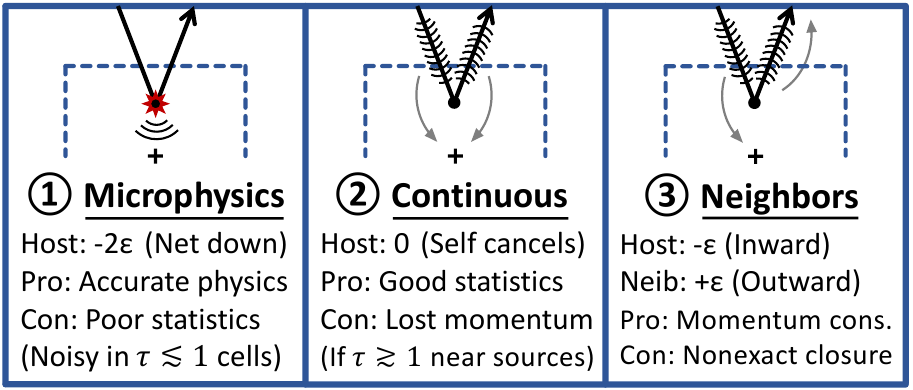}
\includegraphics[width=\columnwidth]{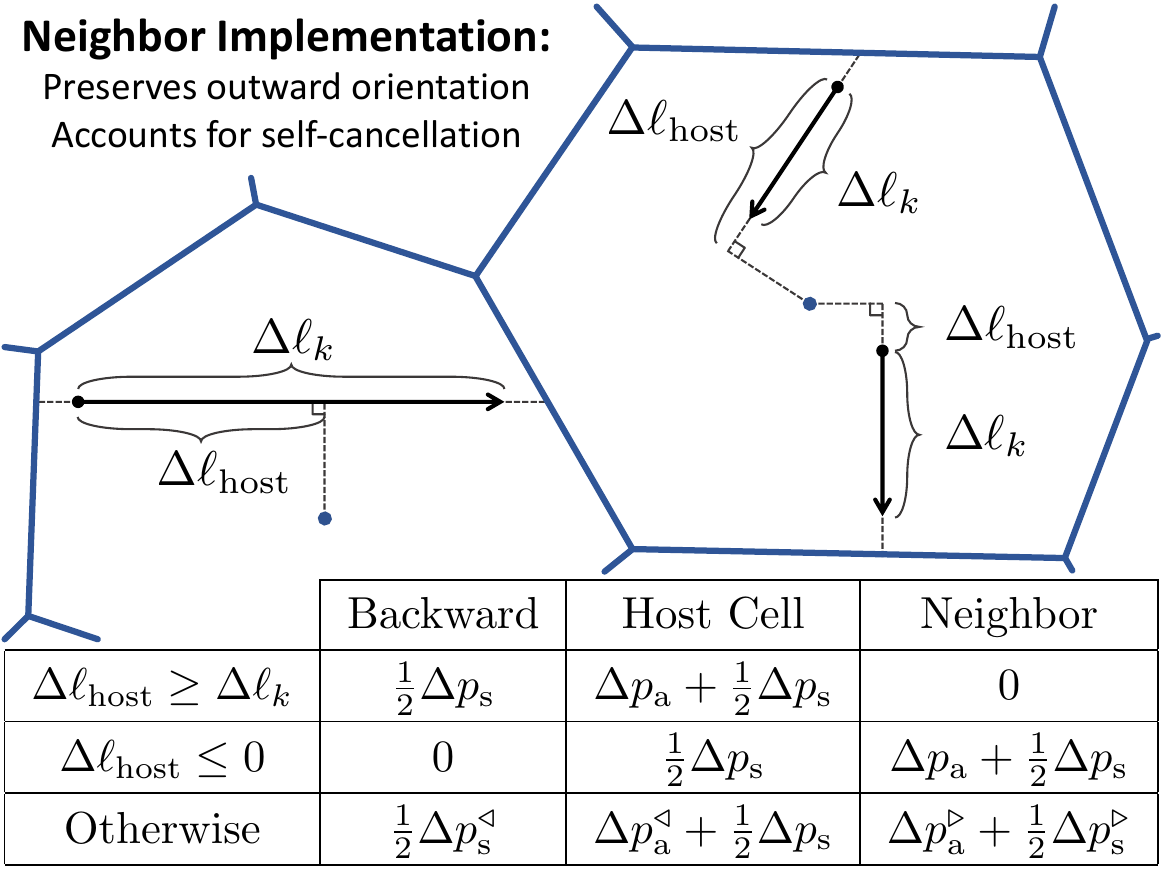}
\vspace{-0.25cm}
\caption{Schematic diagram to motivate the optional neighbor momentum coupling scheme with brief pros and cons of each method. The accompanying diagram illustrates the implementation, which interpolates between volume- and face-integrated methods. First, if $\Delta \ell_\text{host} \geq \Delta \ell_k$, then the momentum is shared with the backward neighbor. Second, if $\Delta \ell_\text{host} \leq 0$, then the momentum is shared with the forward neighbor. Otherwise, the total momentum is split between the host and both neighbors according to the inward $\Delta p^{\triangleleft}$ and outward $\Delta p^{\triangleright}$ oriented portions as described in the text.} \label{fig:momentum_diagram}
\end{figure}

On the other hand, if a similar source is located near a cell interface then a purely neighbor-based method encounters a scenario in which momentum is conserved but imparted to the wrong cells. This is because the cells hosting the photons are transparent to the momentum, even though this is where the unresolved absorption and scattering actually takes place. One solution is to construct a local tessellation around each source as is done by \citet{Hopkins2019}, but this is not practical for MCRT. Instead we propose a scheme which interpolates between the volume- and face-integrated methods, such that the total momentum is split between the host and neighbor cells. Specifically, in highly diffusive regions we expect back-scattering of photons to cancel out most momentum depositions, consistent with a force multiplier of $\sim \tau$ despite having $\sim \tau^2$ scattering events. To reproduce this physical self-cancellation and ensure the momentum flux is conserved but small compared to the energy content, we apply half of the scattering momentum to the host and half to the neighbor. Note that momentum imparted from absorbed photons can safely be applied directly to the forward neighbor. We enforce positive orientation by comparing the path to the center of mass $\bmath{r}_\text{host}$, corresponding to a distance of
\begin{equation} \label{eq:dl_host}
  \Delta \ell_\text{host} \equiv \bmath{n}_k \bmath{\cdot} (\bmath{r}_\text{host} - \bmath{r}_k) \, .
\end{equation}
There are three cases: (i) if $\Delta \ell_\text{host} \geq \Delta \ell_k$ then the momentum is shared with the backward neighbor; (ii) if $\Delta \ell_\text{host} \leq 0$ then the momentum is shared with the forward neighbor; and (iii) otherwise the momentum from equation~(\ref{eq:momentum}) is split between the host and both neighbors according to the inward $\Delta p^{\triangleleft}$ and outward $\Delta p^{\triangleright}$ oriented portions\footnote{Equation~(\ref{eq:momentum}) satisfies the additive property that the total momentum can be separated into disjoint parts: $\Delta p = \Delta p^{\triangleleft} + \Delta p^{\triangleright}$. In practice, we compute the inward oriented absorption portion as $\Delta p_\text{a}^{\triangleleft} = \frac{\varepsilon_k}{c} (1 - e^{-\tau_\text{a,host}})$ where $\tau_\text{a,host} = k_\text{a} \Delta \ell_\text{host}$, the scattering portion as $\Delta p_\text{s}^{\triangleleft} = \Delta p_\text{a}^{\triangleleft} k_\text{s} / k_\text{a}$, and all momenta are applied in the $\bmath{n}_k$ direction. Finally, we note that the backward neighbor corresponds to the shortest face distance in the $-\bmath{n}_k$ direction and can be found when searching for the forward neighbor.}. For clarity, in Figure~\ref{fig:momentum_diagram} we provide a diagram illustrating examples of each of these cases along with a concise implementation table. We also include a schematic diagram of the basic ideas motivating the design of the neighbor momentum coupling scheme.

\newpage

\subsection{Adaptive convergence}
Noise is an inherent feature of MCRT, which can compromise the simulation accuracy when coupling to the hydrodynamics if convergence is not reached. Fortunately, the error is straightforward to quantify because the photon discretization is essentially a Poisson process and the relative signal-to-noise ratio increases with the number of photon packets as $\text{SNR} \propto \sqrt{N}$. For path-based Monte Carlo estimators the distribution for continuous energy deposition $\Delta E$ (from equations \ref{eq:energy_deposition} and \ref{eq:residence_energy}) can be highly unpredictable. However, if we estimate the discretized path integration as a weighted Poisson process then the relative error in cell $i$ is given by\footnote{In this subsection we abbreviate some notation to simplify the discussion of statistical moments, i.e. we do not explicitly include repetitive cell and path subscripts such as $(\sum_\text{paths} \Delta E)_i$.}
\begin{equation} \label{eq:relative_error}
  \delta_i \equiv \frac{\sqrt{\sum \Delta E^2}}{\sum \Delta E} \, .
\end{equation}
In practice, there are also a fixed number of pre-existing packets carried over from previous time steps. These contribute to the overall path statistics but their variance contribution cannot be reduced. Our approach is to reweight equation~(\ref{eq:relative_error}) by the appropriate relative energy, i.e. $\sum \Delta E / (\sum \Delta E + \sum \Delta E_\text{pre})$. This is equivalent to ignoring the pre-existing variance in on-the-fly estimates, but properly accounts for adaptive convergence in time-dependent simulations. We note that the property of diminishing returns can necessitate a large number of photons for high-resolution three-dimensional simulations. However, with enough computational resources the noise can always be maintained below a specified tolerance level, e.g. $\delta_\text{goal} = 0.1$ for smaller than ten per cent local error.

Additionally, the global number of photons required to achieve convergence strongly depends on the overall conditions of the gas and radiation. Therefore, we implement an adaptive convergence scheme such that the emission and propagation of new photons during each time step is performed iteratively in batches\footnote{Energy quantities depend on the cumulative number of photon packets. Thus, to ensure consistency a minor memory cost is associated with adaptive convergence to store both the previous and current deposition arrays.}. We employ a threshold-based metric to ensure the unconverged fraction remains low, e.g. $f_\text{goal} = 0.1$ for over ninety per cent confidence of global convergence. Specifically, the effective fraction is
\begin{equation}
  f \equiv \sum_{\delta_i > \delta_\text{goal}} w_i \, ,
\end{equation}
where the sum is over all unconverged cells, i.e. $\delta_i > \delta_\text{goal}$, and the cell weight is proportional to the new energy density from the time step $w_i \propto \sum \Delta E / V_i$ normalized such that $\sum w_i = 1$. The size of subsequent batches is based on the current level of convergence. In our current implementation we employ a rapid growth mode to find the order of magnitude for the number of photons needed for convergence and a slower reduction mode until the threshold is attained, i.e. $f < f_\text{goal}$. If this is not satisfied then to avoid overcompensating during initial iterations we multiply the unconverged fraction by a factor of $1 - \Delta N_\text{ph} / 2 N_\text{ph}$, where $\Delta N_\text{ph}$ is the number of photons in the most recent batch and $N_\text{ph}$ is the cumulative total from all batches. If $f > 1/2$ the growth mode increases the number of photons in the next batch by a factor of $2^{2f-1}$, otherwise the reduction mode decreases the next batch by a factor of $1.1^{-\log(2 f) / \log(2 f_\text{goal})}$. For additional control we also limit the batch sizes between minimum and maximum values such that $\Delta N_\text{ph,min} < \Delta N_\text{ph} < \Delta N_\text{ph,max}$.

Finally, we note that the variance-to-mean ratio may lead to artificial convergence if the contributions are from highly skewed distributions. In this case it is possible to probe higher moments, such as the variance of the variance (VOV), which measures the relative statistical uncertainty in the estimated relative error and can be approximated as $\text{VOV} = \sum (\Delta E - \sum \Delta E)^4 / \sum (\Delta E - \sum \Delta E)^2$. Such statistics may still not provide the full picture but can be easily calculated on the fly and included in output files as a way to intelligently lower resolution to ensure sufficiently high signal to noise for internal and observed quantities. Still, MCRT convergence schemes may also benefit from tests that are not based on extrapolation. For example, one might consider the variance between chains of photon trajectories as is done with the Gelman--Rubin diagnostic widely employed in Bayesian inference \citep{GelmanRubin1992}. We expect such tests to be more robust but also difficult to implement in practical applications so we leave this for future studies.

\subsection{Discrete-Diffusion Monte Carlo}
\label{sec:DDMC}
When the mean free paths of photons are unresolved in a simulation setup those cells are within the radiative diffusion regime. In this case the transport term can be approximated by an isotropic diffusion process. Specifically, we apply Fick's law as a closure relation to the zeroth order moment equation, such that the radiative flux is proportional to the energy density gradient $\bmath{F} = -c \bmath{\nabla} u / 3k_\text{s}$. The basic form of the RTE without source terms is now\footnote{Under the continuous absorption method only scattering is included as photons are reweighted according to the $e^{-\tau_\text{a}}$ correction factor.}
\begin{equation} \label{eq:diffusion_equation}
  \frac{1}{c} \frac{\partial u}{\partial t} = \bmath{\nabla} \bmath{\cdot} \left( \frac{\bmath{\nabla}u}{3 k_\text{s}} \right) \equiv \mathcal{L}u \, .
\end{equation}
Since \Arepo\ is a finite-volume code based on a Voronoi tessellation of mesh generating points, we recast the local linear operator on the right-hand side of equation~(\ref{eq:diffusion_equation}) into the form \citep[for a similar discussion including anisotropic diffusion see][]{Kannan2016}
\begin{equation}
  \mathcal{L}u = \lim_{V\rightarrow0} \frac{1}{V} \int \bmath{\nabla} \bmath{\cdot} \left( \frac{\bmath{\nabla}u}{3 k_\text{s}} \right)\,\text{d}V \, ,
\end{equation}
and apply Gauss's divergence theorem to get
\begin{equation}
  \mathcal{L}u = \lim_{V\rightarrow0} \frac{1}{V} \oint \frac{\bmath{\nabla}u}{3 k_\text{s}} \bmath{\cdot} \text{d}\bmath{A} \, .
\end{equation}
Therefore, the discretized radiation energy density in a finite-volume scheme on an unstructured mesh for each cell $i$ over all neighbor cells $\delta i$ is
\begin{equation} \label{eq:k_leak}
  \mathcal{L}u_i = \sum_{\delta i} \frac{A_{\delta i}}{V_i} \frac{(u_{\delta i} - u_i)}{3 \Delta \tau_{\text{s},\delta i}} \equiv \sum_{\delta i} k_\text{leak}^{\delta i} (u_{\delta i} - u_i) \, ,
\end{equation}
where $V_i$ is the current cell volume, $A_{\delta i}$ is the area of the shared face, and $\Delta \tau_{\text{s},\delta i} \equiv (k_{\text{s},i} + k_{\text{s},\delta i}) \Delta r_{\delta i} / 2$ is the optical depth between the two mesh generating points, i.e. $\Delta r_{\delta i} \equiv \| \bmath{r}_{\delta i} - \bmath{r}_i \|$ with the cell interface halfway between\footnote{In steady-state radiative equilibrium the flux is $\bmath{F}_\nu = -\frac{4 \pi}{3 k_\nu} \frac{\text{d}B_\nu}{\text{d}T} \bmath{\nabla}T$, which implies that the DDMC leakage coefficients should be constructed from the Rosseland mean: $k_\text{R} \equiv \int_0^\infty \frac{\text{d}B_\nu}{\text{d}T}\,\text{d}\nu / \int_0^\infty k_\nu^{-1} \frac{\text{d}B_\nu}{\text{d}T}\,\text{d}\nu$.}. We note that the general forms for the `leakage coefficients' $k_\text{leak}^{\delta i}$ in equation~(\ref{eq:k_leak}) reduce to the expressions previously found for non-uniform Cartesian and spherical geometries \citep[e.g.][]{Densmore2007,Abdikamalov2012,Tsang2018}. In the MCRT interpretation this discretization of the diffusion operator provides the mechanism for spatial transport of photon packets, with the final form of equation~(\ref{eq:k_leak}) arranged to highlight photon flux conservation across cell interfaces.

\subsubsection{Semi-deterministic DDMC momentum deposition}
The MC procedure and RHD coupling are similar to those of continuous MCRT. However, the optical depth to scattering and distance to the neighboring cell are replaced by an effective distance to leakage drawn from an exponential distribution, i.e. $\Delta \ell_k = -\ln \zeta / k_\text{leak}^{\delta i}$. The traversed optical depth is then $\tau = k \Delta \ell_k$, such that the energy deposition and residence energy density are still given by equations~(\ref{eq:energy_deposition}) and (\ref{eq:residence_energy}), respectively. However, the typical momentum imparted according to equation~(\ref{eq:momentum}) is too large by a factor of $\tau_{\text{s},\delta i}$, which can be seen via direct substitution of the leakage coefficient: $\langle \Delta p_\text{tot} \rangle \approx \tau \varepsilon_k / c = k \varepsilon_k / c k_\text{leak}^{\delta i} = 3 k \Delta \tau_{\text{s},\delta i} V_i \varepsilon_k / c A_{\delta i} \approx \Delta \tau_{\text{s},\delta i} \tau_i \varepsilon_k / c$, with the final approximation valid for a spherical cell with radius $r_i$ and optical depth $\tau_i = k r_i$. Following after equation~(\ref{eq:momentum_rate}) but including both scattering and absorption gives
\begin{equation} \label{eq:DDMC-momentum}
  \Delta \dot{\bmath{p}} = \int \frac{k \bmath{F}}{c} \text{d}V = \oint \frac{u}{3} \text{d}\bmath{A} \approx \sum_{\delta i} \frac{\bmath{A}_{\delta i}}{3} \bar{u} \, ,
\end{equation}
where the bar denotes the average value at the cell interface, which is approximately $\bar{u} \equiv (u_i + u_{\delta i} k_i / k_{\delta i})/2$ for an unstructured mesh and also accounts for a changing absorption coefficient across neighbors. To our best knowledge, this semi-deterministic DDMC momentum scheme has not appeared previously in the literature. It is variance reducing and efficiently applied at the end of the MCRT calculations.

\subsubsection{Hybrid IMC--DDMC}
\label{sec:hybrid-DDMC}
The DDMC method is accurate as long as the diffusion approximation holds within the host cell, which can be violated when transitioning to optically thin regions. Therefore, following \citet{Densmore2007} we implemented a hybrid transport scheme in which DDMC packets can convert to spatially continuous MC packets and vice versa, depending on whether the cell optical depth is sufficiently large, i.e. how $\tau_i = k_{\text{s},i} \min\{\Delta r_{\delta i}\}$ compares to $\tau_\text{DDMC}$. We briefly summarize the main ideas of the hybrid scheme adopting the `asymptotic diffusion limit' as the interfacing boundary condition. If $\tau_i < \tau_\text{DDMC}$, then the leakage coefficient is redefined to be
\begin{equation}
  k_{\text{DDMC} \rightarrow \text{MC}}^{\delta i} = \frac{1}{3 \Delta r_{\delta i}} \frac{2}{k_{\text{s},i} \Delta r_{\delta i} + 2 \lambda} \, ,
\end{equation}
where $\lambda \approx 0.7104$ is the constant extrapolation distance \citep{Habetler1975}. If this corresponds to the minimum distance then the DDMC packet becomes an MC packet with a random position on the cell interface and an isotropic outward direction. On the other hand, if an MC packet moves into a neighboring cell that is optically thick then it is converted into a DDMC packet in that cell with probability
\begin{equation} \label{eq:DDMC-prob}
  P_{\text{MC} \rightarrow \text{DDMC}}^{\delta i} = \frac{2}{k_{\text{s},\delta i} \Delta r_{\delta i} + 2 \lambda} \left(\frac{2}{3} + \mu \right) \, ,
\end{equation}
where $\mu$ is the directional cosine for the MC packet with respect to the cell interface. Otherwise, the packet scatters back into the original cell with a random isotropic inward direction. To have a valid probabilistic interpretation, we require $P_{\text{MC} \rightarrow \text{DDMC}}^{\delta i} \in [0,1]$ for $\mu \in (0,1]$, which imposes a condition that $\tau_\text{DDMC} \gtrsim 2$, although we suggest a more conservative choice of $\tau_\text{DDMC} = 5$. To further explore this hybrid scheme, in Appendix~\ref{sec:DDMC_BCs} we demonstrate the validity of our new semi-deterministic DDMC momentum scheme from equation~(\ref{eq:DDMC-momentum}) across extreme DDMC--MC transitions.

\subsection{DDMC with advection}
\label{sec:DDMC_advection}
We now present a new unsplit DDMC scheme to incorporate an advection term into the leakage coefficients, which can be important in the dynamical diffusion regime where $\tau \bmath{v} / c \gtrsim 1$. Following Section~\ref{sec:DDMC}, the basic form of the RTE without source terms is now
\begin{equation} \label{eq:diffusion_equation_advection}
  \frac{1}{c} \frac{\partial u}{\partial t} = \bmath{\nabla} \bmath{\cdot} \left( \frac{\bmath{\nabla}u}{3 k_\text{s}} - \frac{\bmath{v} u}{c} \right) \equiv \mathcal{L}u \, .
\end{equation}
In the finite-volume framework, we recast the local linear operator on the right-hand side of equation~(\ref{eq:diffusion_equation_advection}) into the form
\begin{equation}
  \mathcal{L}u = \lim_{V\rightarrow0} \frac{1}{V} \int \bmath{\nabla} \bmath{\cdot} \left( \frac{\bmath{\nabla}u}{3 k_\text{s}} - \frac{\bmath{v} u}{c} \right)\,\text{d}V \, ,
\end{equation}
and apply Gauss's divergence theorem to get
\begin{equation}
  \mathcal{L}u = \lim_{V\rightarrow0} \frac{1}{V} \oint \left( \frac{\bmath{\nabla}u}{3 k_\text{s}} - \frac{\bmath{v} u}{c} \right) \bmath{\cdot} \text{d}\bmath{A} \, .
\end{equation}
Therefore, the discretized radiation energy density in a finite-volume scheme for each cell $i$ over all neighbor cells $\delta i$ is
\begin{align} \label{eq:k_leak_advection}
  &\mathcal{L}u_i = \sum_{\delta i} \frac{A_{\delta i}}{V_i} \left[ \frac{u_{\delta i} - u_i}{3 \Delta \tau_{\text{s},\delta i}} - \frac{\bar{v} \bar{u}}{c} \right] \\
  &= \sum_{\delta i} \frac{A_{\delta i}}{V_i} \left[ \left( \frac{1}{3 \Delta \tau_{\text{s}, \delta i}} - \frac{\bar{v}}{2c} \right) u_{\delta i} - \left( \frac{1}{3 \Delta \tau_{\text{s}, \delta i}} + \frac{\bar{v}}{2c} \right) u_i \right] \, , \notag
\end{align}
where $V_i$ is the current cell volume, $A_{\delta i}$ is the area of the shared face, and $\Delta \tau_{\text{s},\delta i} \equiv (k_{\text{s},i} + k_{\text{s},\delta i}) \Delta r_{\delta i} / 2$ is the optical depth between the two mesh generating points, i.e. $\Delta r_{\delta i} \equiv \| \bmath{r}_{\delta i} - \bmath{r}_i \|$ with the cell interface halfway between. The bar denotes the average at the cell interface, which to first order is approximately $\bar{u} \equiv (u_i + u_{\delta i}) / 2$ and $\bar{v} \equiv (\bmath{v}_i + \bmath{v}_{\delta i}) \bmath{\cdot} (\bmath{r}_{\delta i} - \bmath{r}_i) / 2 \Delta r_{\delta i}$. In the MCRT interpretation this discretization provides the mechanism for spatial transport of photon packets, with the final form of equation~(\ref{eq:k_leak_advection}) arranged to highlight the asymmetric leakage due to the preferred direction of the gas motion. In fact, the inhibited- or enhanced-leakage coefficients are modified to reflect advective transport, which can be succinctly implemented by noticing that $k_\text{leak,adv}^{\delta i} = k_\text{leak}^{\delta i} (1 + 3 \Delta\tau_{\text{s},\delta i} \bar{v} / 2 c)$.

We note that negative leakage coefficients are possible when the oncoming flow overwhelms the probability of upstream diffusion, and should be ignored. While this interpretation is physically meaningful, such a logical inconsistency is indicative of either insufficient spatial resolution or that a fully relativistic treatment of radiative transfer is necessary. Of course, even first-order comoving-frame DDMC is not needed for our present applications, but we hope this will serve as a useful tool in the DDMC community. The Doppler correction terms may also be treated in an analogous fashion. Finally, the momentum should also be modified:
\begin{align}
  \Delta \dot{\bmath{p}}_\text{adv} &= \int \bmath{\nabla} \bmath{\cdot} \left( \frac{\bmath{v} \otimes \bmath{F}}{c^2} \right) \text{d}V \notag \\
  &= - \oint \left( \bmath{v} \otimes \frac{\bmath{\nabla}u}{3 c k} \right) \bmath{\cdot} \text{d}\bmath{A} \notag \\
  &\approx -\sum_{\delta i} \frac{A_{\delta i} \bar{\bmath{v}}}{3 c \Delta \tau_{\delta i}} (u_{\delta i} - u_i) \, ,
\end{align}
such that the momentum correction over the time step is
\begin{equation}
  \Delta \bmath{p}_\text{adv} \approx -\sum_{\delta i} k_\text{leak}^{\delta i} V_i \Delta t (u_{\delta i} - u_i) \frac{\bar{\bmath{v}}}{c} \, .
\end{equation}

\begin{figure}
\centering
\includegraphics[width=\columnwidth]{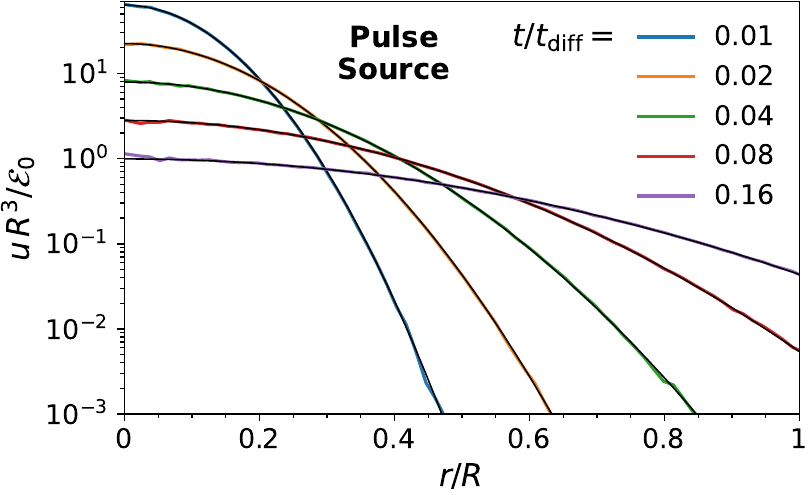}
\caption{Radiation energy density $u(r)$ for a pulse source diffusing in a uniform medium over several doubling times, $t = \{1,2,4,8,16\} \times 10^{-2}\,t_\text{diff}$, where the diffusion time is $t_\text{diff} = \frac{3}{2} \tau t_\text{light} = 3 k R^2 / 2 c$. The analytic solution from equation~(\ref{eq:grey_diffusion_pulse}) is shown by the black curves. For convenience the axes have been rescaled into dimensionless units.} \label{fig:grey_diffusion_pulse}
\end{figure}

\begin{figure}
\centering
\includegraphics[width=\columnwidth]{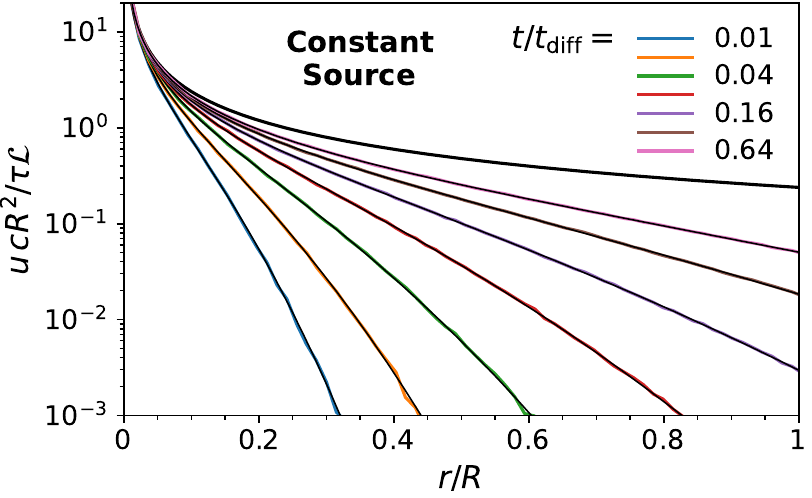}
\caption{Radiation energy density $u(r)$ for a constant source diffusing in a uniform medium over several doubling times, $t = \{1,2,4,8,16,32,64\} \times 10^{-2}\,t_\text{diff}$, where the diffusion time is $t_\text{diff} = \frac{3}{2} \tau t_\text{light} = 3 k R^2 / 2 c$. The analytic solution from equation~(\ref{eq:grey_diffusion_const}) is shown by the black curves with the final thick black curve showing the steady-state solution, $u = 3 k \mathcal{L} / 4 \pi c r$.} \label{fig:grey_diffusion_const}
\end{figure}

\section{Test Problems}
\label{sec:tests}

\subsection{Gray diffusion}
\label{sec:grey_diffusion}
To test the spatial transport of the MC particles we consider pure scattering in optically thick media. In the case of a constant scattering coefficient this is equivalent to a random walk with a mean free path of $\lambda_\text{mfp} = k^{-1}$. For an arbitrary reference length scale~$R \gg \lambda_\text{mfp}$, the light-crossing and diffusion times are $t_\text{light} = R/c$ and $t_\text{diff} = \frac{3}{2} \tau t_\text{light} = 3 k R^2 / 2 c$. Therefore, the evolution of the radiation energy density is governed by a diffusion equation $\partial u / \partial t = (c/3k) \nabla^2 u$ and the solution given an initial point source impulse of energy $\mathcal{E}_0$ is
\begin{equation} \label{eq:grey_diffusion_pulse}
  \tilde{u} = \frac{e^{-\tilde{r}^2/2\tilde{t}}}{(2 \pi \tilde{t})^{3/2}} \, ,
\end{equation}
where we have rescaled into dimensionless units with radius $\tilde{r} = r / R$, time $\tilde{t} = t / t_\text{diff}$, and energy density $\tilde{u} = u\,R^3/\mathcal{E}_0$.

Our spatial transport test consists of a low-resolution three-dimensional Cartesian grid initialized with $10^6$ photon packets at the center at $t = 0$. The mesh configuration and resolution do not matter for this test because we output the photon packets and bin their positions in spherical shells for statistics that directly correspond to the analytic solution from equation~(\ref{eq:grey_diffusion_pulse}). We have verified that the cell-based energy density and momentum estimators give the same results, although path-based estimators represent averages over discrete time steps and cell volumes. Fig.~\ref{fig:grey_diffusion_pulse} shows the radiation energy density radial profile over several doubling times, $t = \{1,2,4,8,16\} \times 10^{-2}\,t_\text{diff}$. The simulation provides excellent agreement with the exact analytical solution of equation~(\ref{eq:grey_diffusion_pulse}). To simulate an infinite domain, we employ periodic boundary conditions but keep track of the domain tiling to retain the absolute positions of each photon packet.

\begin{figure}
\centering
\includegraphics[width=\columnwidth]{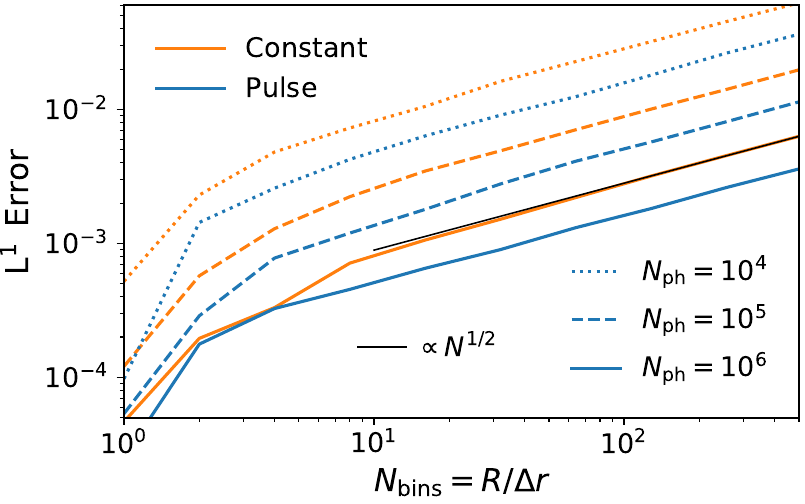}
\caption{Volume-weighted $L^1$ error of the radiation energy density, $\int |u - u_\text{exact}|\,\text{d}V / \int \text{d}V$, for the pulse and constant source diffusion tests as a function of the number of radial bins $N_\text{bins}$. We also show results with different numbers of photon packets $N_\text{ph} \in \{10^4, 10^5, 10^6\}$ (per diffusion timescale for the constant source case). The errors are normalized to the total radiation energy and time averaged. The MCRT noise follows the expected $\propto \sqrt{N_\text{bins} / N_\text{ph}}$ relation for the number of bins and photon packets.} \label{fig:grey-diffusion-L1}
\end{figure}

\begin{figure}
\centering
\includegraphics[width=\columnwidth]{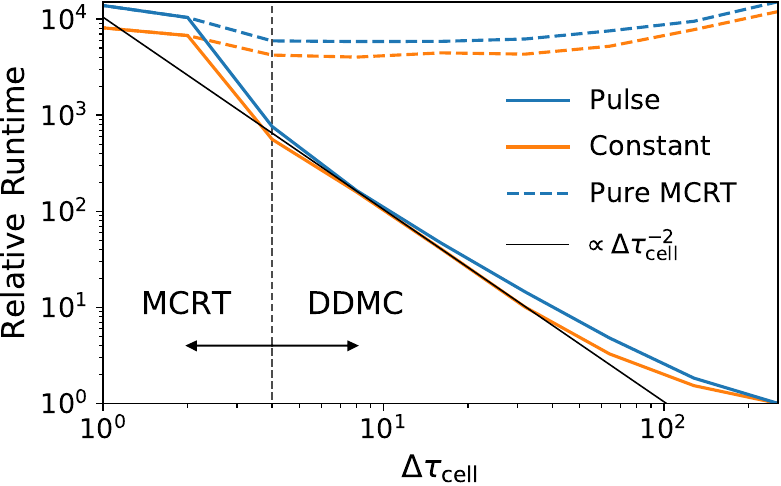}
\caption{Relative simulation runtime for the pulse and constant source diffusion tests as a function of the cell optical depth resolution $\Delta \tau_\text{cell}$. Aside from code overheads, these tests demonstrate that pure MCRT is approximately independent of the simulation resolution, while hybrid MCRT--DDMC exhibits the expected $\propto \Delta \tau_\text{cell}^{-2}$ speedup from bypassing sub-grid scattering calculations. The simulations have the same characteristic optical depth of $\tau = 512$ and are run for a full diffusion timescale on a periodic domain.} \label{fig:DDMC-scaling}
\end{figure}

We also test the spatial diffusion of photon packets under a constant luminosity point source $\mathcal{L}$. This is the same setup as before but the evolution of the radiation energy density is given by
\begin{equation} \label{eq:grey_diffusion_const}
  \tilde{u} = \frac{3}{4 \pi \tilde{r}} \, \text{erfc}\left(\frac{\tilde{r}}{\sqrt{2 \tilde{t}}}\right) \, ,
\end{equation}
where we have again rescaled into dimensionless units with radius $\tilde{r} = r / R$, time $\tilde{t} = t / t_\text{diff}$, energy density $\tilde{u} = u\,c R^2 / \tau \mathcal{L}$, and the steady-state solution is $\tilde{u}|_{t\rightarrow\infty} = 3 / 4 \pi \tilde{r}$. For this test we emit photon packets from the center at a constant rate of $5 \times 10^6 / t_\text{diff}$. Fig.~\ref{fig:grey_diffusion_const} shows the radiation energy density over several doubling times, $t = \{1,2,4,8,16,32,64\} \times 10^{-2}\,t_\text{diff}$. The simulation provides excellent agreement with the analytical solution of equation~(\ref{eq:grey_diffusion_const}).

With exact solutions in hand we can directly quantify the numerical error of scattering-dominated transport. In Fig.~\ref{fig:grey-diffusion-L1} we present the volume-weighted $L^1$ error of the radiation energy density, $\int |u - u_\text{exact}|\,\text{d}V / \int \text{d}V$, for the pulse and constant source diffusion tests as a function of the number of radial bins $N_\text{bins}$ and photon packets $N_\text{ph}$. The errors are normalized such that the total radiation energy is one, even for the constant source that would otherwise grow in time. We are also careful to integrate the analytic solutions from equations~(\ref{eq:grey_diffusion_pulse}) and (\ref{eq:grey_diffusion_const}) over the same volumes as the MCRT radial bins. The time-averaged comparisons therefore represent the precise numerical error due to the random walk process. In fact, we recover the expected $\propto \sqrt{N_\text{bins} / N_\text{ph}}$ noise relations for the number of bins and photon packets. We note that path-based estimators would allow photons to contribute to multiple cells thereby significantly reducing the error beyond what is shown. However, a fair comparison in general three-dimensional geometry would require a more complex treatment of cell volumes and integrating over the lagged path time steps. These simple tests stress the need for on-the-fly convergence criteria in MCRT RHD simulations, where the resolution is largely determined by the gas dynamics.

Finally, we demonstrate the speedup that the DDMC scheme provides for scattering-dominated transport. In Fig.~\ref{fig:DDMC-scaling} we present the relative runtime for the pulse and constant source diffusion tests as a function of the cell optical depth resolution $\Delta \tau_\text{cell}$. The simulations have a total characteristic optical depth of $\tau = 512$ and are run for a full diffusion timescale on a periodic domain, so pure MCRT has $\gtrsim 10^5$ scattering events per photon. The runtimes are scaled to the fastest DDMC timings ($\approx 1$ second on a laptop computer), which employ a large number of photon packets ($\approx 10^7$). The MCRT timings are approximately constant while the DDMC speedup follows the expected $\propto \Delta \tau_\text{cell}^{-2}$ scaling. Deviations are due to various overheads and nuances related to running this simple test while including redundant physics in an unstructured mesh code.

\begin{figure}
\centering
\includegraphics[width=\columnwidth]{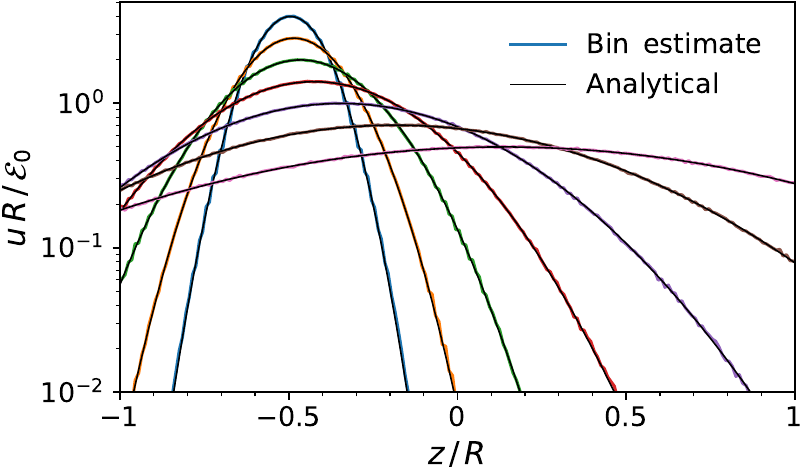}
\caption{Radiation energy density $u(z)$ for gray diffusion undergoing constant relative motion in a uniform medium over several doubling times, $t = \{1,2,4,8,16,32,64\} \times 10^{-2}\,t_\text{diff}$, where the diffusion time is $t_\text{diff} = \frac{1}{2} \tau t_\text{light} = k R^2 / 2 c$. The numerical solution (colored curves) employ the DDMC advection scheme and the analytic solution (black curves) are from equation~(\ref{eq:grey_diffusion_pulse}) but shifted in time. For convenience the axes are in dimensionless units.} \label{fig:grey_advection}
\end{figure}

\subsection{Diffusion with advection}
We now provide two basic tests to demonstrate the accuracy of our DDMC advection scheme. In Fig.~\ref{fig:grey_advection} we validate the ability to capture flows from an Eulerian reference frame by considering the same setup as Section~\ref{sec:grey_diffusion} but with a constant velocity. By symmetry the solution is the same as the one-dimensional version of equation~(\ref{eq:grey_diffusion_pulse}) but with a coordinate boost of $z \rightarrow z - v_0 t$. We set the advection crossing time to be equal to the diffusion time, i.e. $v_0 = R / t_\text{diff} = 2 c / \tau$. We find that this new unsplit approach is both highly efficient and accurate.

\begin{figure}
\centering
\includegraphics[width=\columnwidth]{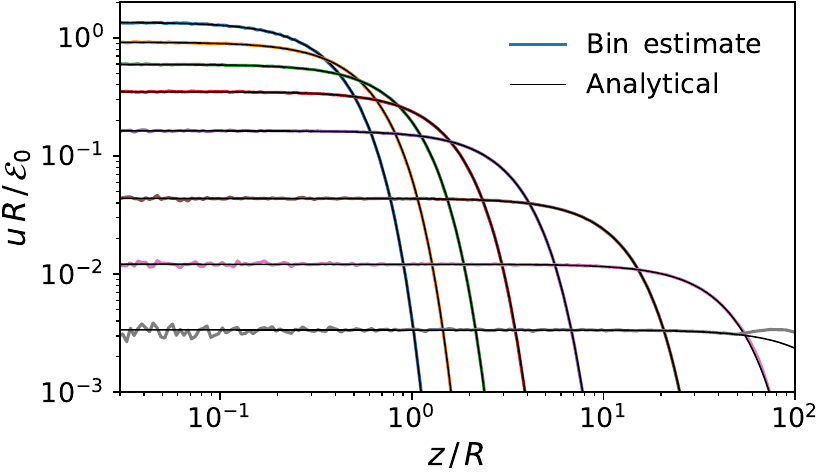}
\caption{Radiation energy density $u(z)$ for gray diffusion in a uniform medium undergoing homologous expansion shown at times of $t = \{1,2,4,8,16,32,48,64\} \times 10^{-2}\,t_\text{diff}$, where the diffusion time is $t_\text{diff} = \frac{1}{2} \tau t_\text{light} = k R^2 / 2 c$. The numerical solution (colored curves) employ the DDMC advection scheme and the analytic solution (black curves) are from equation~(\ref{eq:homologous_advection}). The bump in the final curve is due to an unphysical boundary condition. For convenience the axes are in dimensionless units.} \label{fig:grey_expansion}
\end{figure}

For the second test, we consider the homologous stretching of a one-dimensional infinite plane-parallel slab. In this case the velocity is given by $v(z) = v_0 z / R$, and the evolution of the radiation energy density is governed by the partial differential equation $\partial u / \partial t = (c/k) \partial^2 u / \partial z^2 - (v_0/R) (u + z \partial u / \partial z)$. The solution can be derived with the ansatz that diffusive stretching simply modifies the elapsed time on the global radiation clock. Upon substitution of $\tilde{t} \rightarrow \tilde{s}(\tilde{t})$ we find the ansatz reduces the problem to an ordinary differential equation $\tilde{s}'(\tilde{t}) = 1 + 2 \tilde{v}_0 \tilde{s}(\tilde{t})$ subject to the condition that $\tilde{s}(0) = 0$ (see dimensionless definitions below equation~\ref{eq:homologous_advection}). Therefore, the full solution for an initial point source impulse of energy $\mathcal{E}_0$ undergoing homologous expansion is
\begin{equation} \label{eq:homologous_advection}
  \tilde{u} = \frac{e^{-\tilde{z}^2 / 2 \tilde{s}}}{\sqrt{2\pi \tilde{s}}} \qquad \text{where} \quad \tilde{s} = \frac{e^{2 \tilde{v}_0 \tilde{t}} - 1}{2 \tilde{v}_0} \, .
\end{equation}
We have rescaled into dimensionless units with position $\tilde{z} = z / R$, velocity $\tilde{v}_0 = v_0 k R / 2 c$, time $\tilde{t} = t / t_\text{diff}$, with $t_\text{diff} = k R^2 / 2 c$, and energy density $\tilde{u} = u R / \mathcal{E}_0$. In Fig.~\ref{fig:grey_expansion} we validate the DDMC advection scheme under this more stringent test. Specifically, we set the characteristic expansion timescale to be equal to the diffusion time, i.e. $v_0 = R / t_\text{diff} = 2 c / \tau$.

For completeness, we also provide the analytic solution for diffusion under homologous expansion in spherical geometry. The derivation mirrors that of the one-dimensional slab with the full solution being
\begin{equation} \label{eq:homologous_advection_sphere}
  \tilde{u} = \frac{e^{-\tilde{r}^2 / 2 \tilde{s}}}{(2\pi \tilde{s})^{3/2}} \qquad \text{where} \quad \tilde{s} = \frac{e^{2 \tilde{v}_0 \tilde{t}} - 1}{2 \tilde{v}_0} \, .
\end{equation}
We have rescaled into dimensionless units with radius $\tilde{r} = r / R$, velocity $\tilde{v}_0 = v_0 3 k R / 2 c$, time $\tilde{t} = t / t_\text{diff}$, with $t_\text{diff} = 3 k R^2 / 2 c$, and energy density $\tilde{u} = u R^3 / \mathcal{E}_0$. We can understand this general stretching behavior by considering that a test particle following the Lagrangian flow from homologous expansion has a physical coordinate of $r(t) = r_0 e^{v_0 t / R}$. This can be derived by induction from a Riemann integration of the velocity field with a starting radius $r_0$ and equal time segments $\Delta t = t / N$ such that $r_N = r_0 (1 + v_0 t / R N)^N$, which reduces to the exponential function in the limit as $N \rightarrow \infty$. However, the equation of motion is modified by the random walk process in an invariant manner requiring that $\tilde{s} \approx \tilde{t}$ early on. To the best of our knowledge these homologous diffusion solutions are new and can serve as an additional benchmark for codes wishing to accurately model radiation in the dynamical diffusion regime.

\begin{figure}
\centering
\includegraphics[width=\columnwidth]{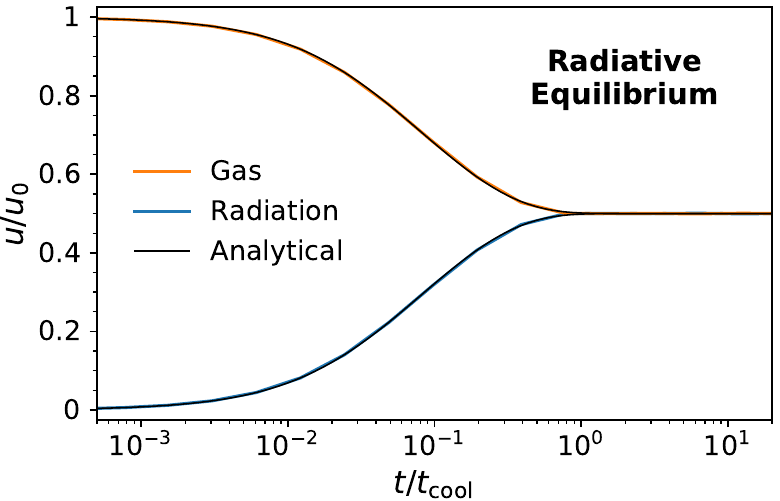}
\caption{Gas and radiation energy densities $u(t)$ to demonstrate the evolution to radiative equilibrium in a uniform medium. Here the cooling time is defined as $t_\text{cool} \equiv 1 / c k_a$. The numerical solution from equation~(\ref{eq:radiative_equilibrium}) is shown by the black curves. For convenience the axes have been rescaled into dimensionless units.} \label{fig:radiative_equilibrium}
\end{figure}

\subsection{Radiative Equilibrium}
\label{sec:radiative_equilibrium}
To test the coupling of the gas internal energy and radiation fields we consider pure absorption in LTE over a uniform medium. Defining $u_r = a_\text{B} T^4$ as in Section~\ref{sec:RT}, the stiff system of equations governing the gas and radiation energy density evolution is
\begin{equation} \label{eq:radiative_equilibrium}
  \frac{\text{d}\tilde{u}_\text{g}}{\text{d}\tilde{t}} = -\frac{\text{d}\tilde{u}}{\text{d}\tilde{t}} = \tilde{u} - \zeta \tilde{u}_\text{g}^4 \, .
\end{equation}
Here we have rescaled the variables such that $\tilde{u}_\text{g} \equiv u_\text{g} / u_0$, $\tilde{u} \equiv u / u_0$, where the initial gas and radiation energy densities are $u_\text{g}|_{t=0} = u_0$ and $u|_{t=0} = 0$. We have also introduced a dimensionless coupling parameter that can be given in terms of the initial temperature and density as $\zeta \equiv a_\text{B} T_0^3 (\gamma-1) \mu / k_\text{B} \rho_0$. For this test we set $\zeta = 8$ so that at equilibrium we have $u|_{t\rightarrow\infty} = u_0 / 2$.

Our radiative equilibrium test consists of a one-zone setup, which converges quickly regardless of the number of photon packets. Fig.~\ref{fig:radiative_equilibrium} shows the time evolution of the mean energy densities, with the MCRT result providing excellent agreement with the exact solution. For this test we start with a small time step and force subsequent time steps to be twice the previous one. We also enabled IMC with an implicitness parameter of $\alpha = 0.5$, which provides higher-order accuracy for this special test case.

\begin{figure}
\centering
\includegraphics[width=\columnwidth]{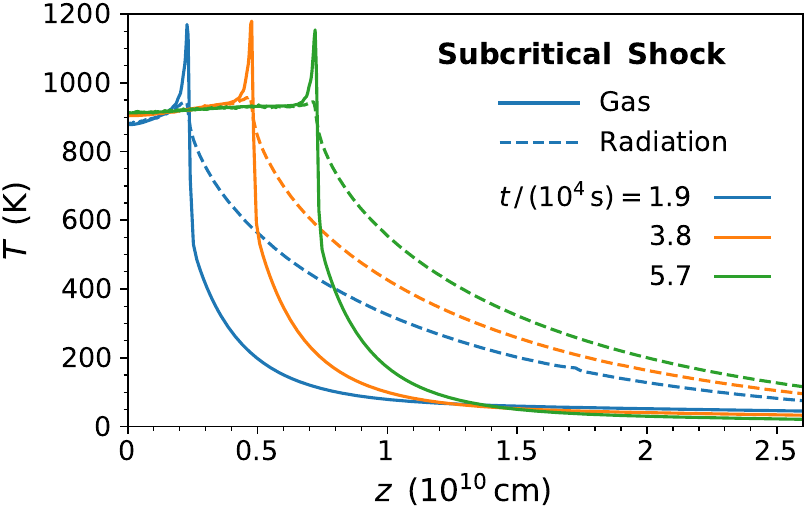}
\caption{Gas and radiation temperatures $T(z)$ for the sub-critical radiative shock test, shown respectively as solid and dashed curves. The initial gas is colliding with a velocity of $v_0 = 6\,\text{km\,s}^{-1}$. The profiles are given at the times $t = \{1.9, 3.8, 5.7\} \times 10^4\,\text{s}$.} \label{fig:radiative_shock_sub}
\end{figure}

\begin{figure}
\centering
\includegraphics[width=\columnwidth]{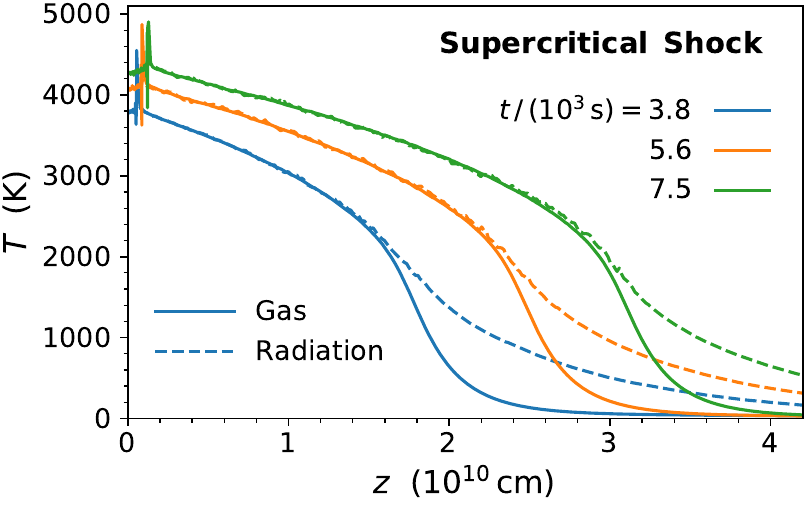}
\caption{Gas and radiation temperatures $T(z)$ for the super-critical radiative shock test, shown respectively as solid and dashed curves. The initial gas is colliding with a velocity of $v_0 = 20\,\text{km\,s}^{-1}$. The profiles are given at the times $t = \{3.75, 5.625, 7.5\} \times 10^4\,\text{s}$.} \label{fig:radiative_shock_super}
\end{figure}

\begin{figure*}
\centering
\includegraphics[width=.99\textwidth]{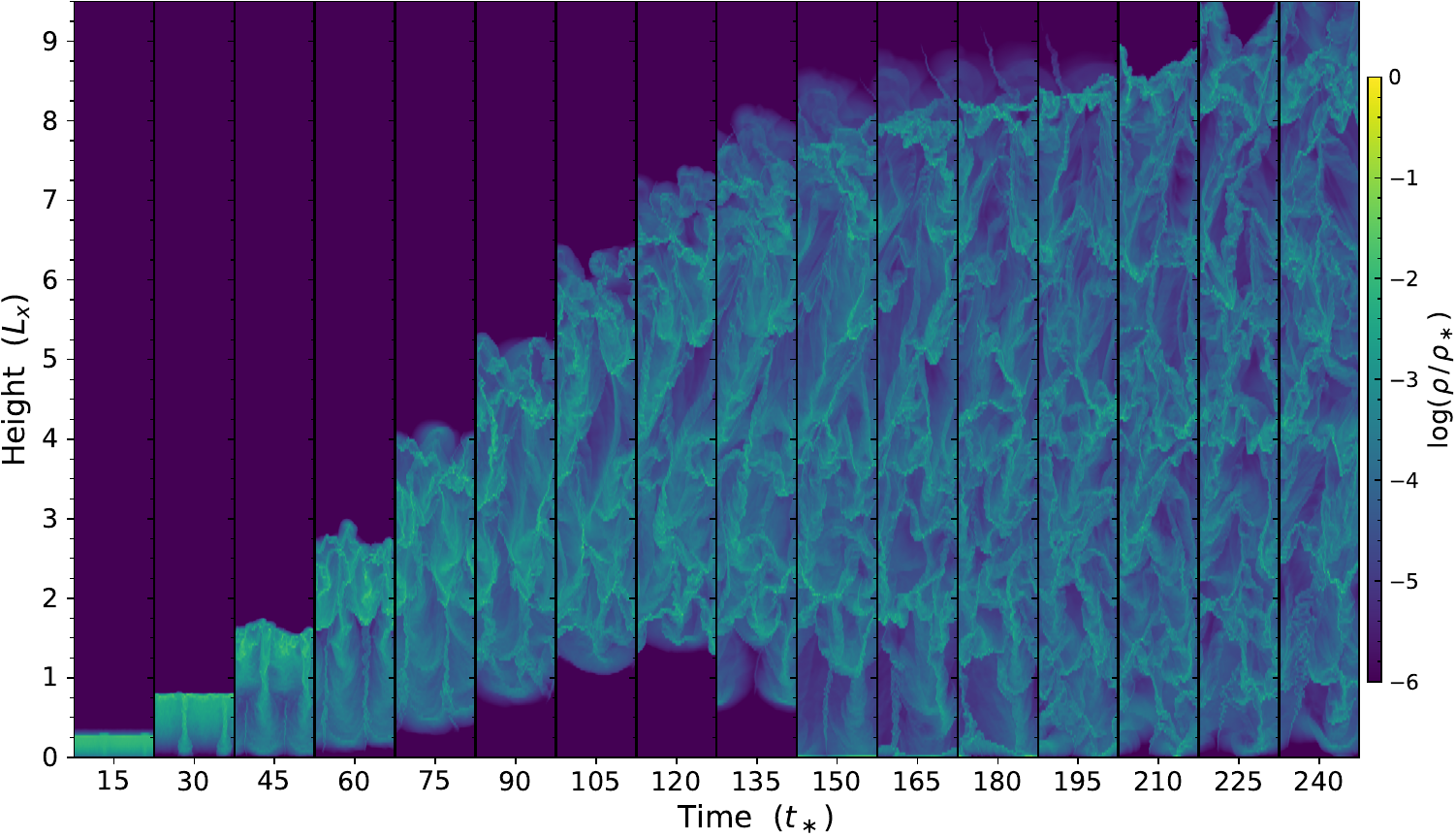}
\caption{Evolution of the normalized gas density $\rho / \rho_\ast$ shown in intervals of $15\,t_\ast$. This view emphasizes the successful launch of the wind despite the gas becoming Rayleigh--Taylor unstable early on. Even after some cold filaments fall back down the gas structure remains highly elongated and turbulent in a quasi-steady state configuration.} \label{fig:levitation_rho}
\end{figure*}

\subsection{Radiative shock}
\label{sec:radiative_shock}
We now test the hydrodynamical coupling by considering the formation of both sub- and super-critical radiative shocks following the initial conditions proposed by \citet{Ensman1994}, which have been reproduced in numerous RHD implementations \citep{Hayes2003,Whitehouse2006,Gonzalez2007,Commercon2011,Noebauer2012,Tsang2015}. We perform the test on a moving mesh starting with a box of radius $R = 7 \times 10^{10}\,\text{cm}$ with uniform density $\rho_0 = 7.78 \times 10^{-10}\,\text{g\,cm}^{-3}$, absorption coefficient $k_\text{a} = 3.115 \times 10^{-10}\,\text{cm}^{-1}$, mean molecular weight $\mu = 1$, adiabatic index $\gamma = 7/5$, and a linear temperature profile decreasing from $T = 85\,\text{K}$ at the center to $T = 10\,\text{K}$ at the edges of the box. We employ IMC with $\alpha = 1$ and allow MC particles to escape at either boundary. The gases on the left and right are colliding toward the center at constant velocity of $v_0 = 6$ and $20\,\text{km\,s}^{-1}$ for the sub- and super-critical shocks, which generates an outwardly propagating shock wave. The thermal radiation diffuses upstream to pre-heat the pre-shock gas to the post-shock temperature \citep{Zeldovich1967}. The temperature profiles for each test are shown in Figs.~\ref{fig:radiative_shock_sub} and \ref{fig:radiative_shock_super} at several different times to illustrate the outward propagation. In both cases, our results are in agreement with previous simulations and analytical studies for the jump conditions across the shock \citep{Mihalas1984}. Further verification of RHD codes could also include the semi-analytic radiative shock solutions of \citep{Lowrie2008}, which are involved in their implementation but provide further insights within the context of gray nonequilibrium radiative diffusion.

\newpage

\begin{figure*}
\centering
\includegraphics[width=.99\textwidth]{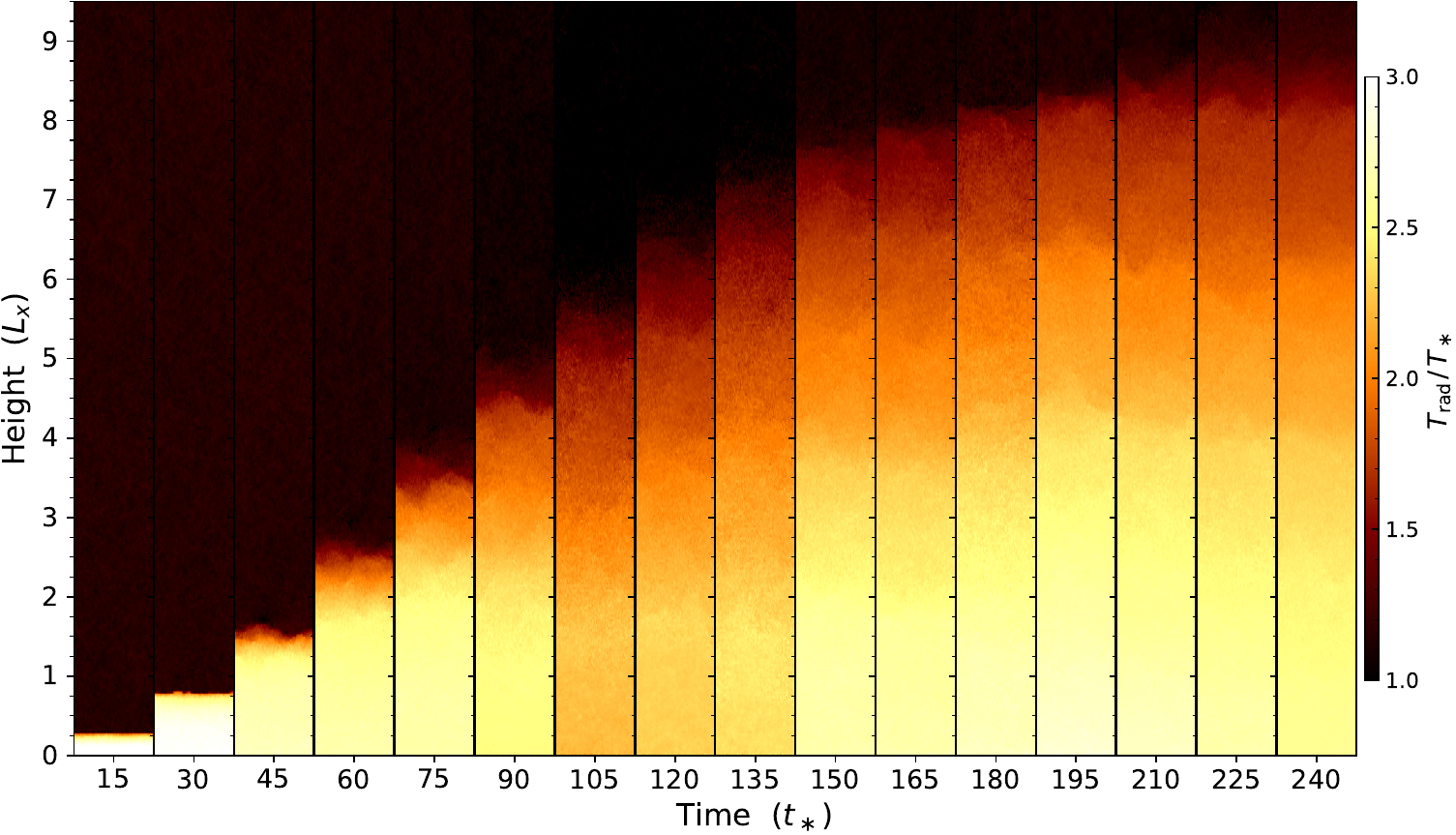}
\caption{Evolution of the normalized radiation temperature $T_\text{r} / T_\ast$, which also closely mirrors the behavior of the gas temperature. The radiation energy is quite smooth due to photon trapping behind the wind front. By $t = 30\,t_\ast$ the radiation efficiently heats and pushes the gas. The rapid expansion, cooling, and escape channels lead to a noticeably reduced temperature by $t = 100\,t_\ast$, but this is built up again by $t = 150\,t_\ast$.} \label{fig:levitation_T_rad}
\end{figure*}

\section{Levitation of optically thick gas}
\label{sec:levitation}
Radiative feedback can play an important role in galaxy formation and evolution by driving supersonic turbulence, reducing the star formation efficiency, and regulating galactic winds \citep[e.g.][]{Thompson2005,Hopkins2011}. In particular, systems with extreme star formation rate densities, including so-called ultraluminous infrared galaxies (ULIRGs), can experience efficient photon trapping as direct ultraviolet (UV) starlight is efficiently reprocessed by dust grains to multiscattered infrared (IR) radiation. In such environments the trapping effect boosts the momentum injection rate by a factor of the optical depth $\tau_\text{IR}$ relative to the intrinsic force budget of $\mathcal{L}/c$. However, in reality the interstellar medium has a hierarchical structure, which facilitates the escape of radiation and the associated momenta through low column density channels. Furthermore, the natural emergence of the Rayleigh--Taylor instability \citep[RTI;][]{Chandrasekhar1961} in the presence of external forces, such as gravity, may limit the coupling of gas and radiation. Modeling these complexities requires multidimensional RHD simulations, which motivated \citet{Krumholz2012} to design a two-dimensional setup to investigate the efficiency of radiation pressure driving of a dusty atmosphere in a vertical gravitational field. Subsequently, several other groups have simulated this levitation setup with different codes and RHD methods, including FLD \citep{Krumholz2012,Davis2014}, VET \citep{Davis2014}, M1 \citep{Rosdahl2015,Kannan2019}, and MCRT \citep{Tsang2015}. As the model and physics are described in detail by each of these authors, we only provide a brief summary along with the \ArepoMCRT\ results for comparison with the previous studies.

\subsection{Simulation setup}
As discussed in \citet{Krumholz2012} and the subsequent studies, the goal is to simulate the evolution of molecular gas at temperatures where dust dominates the opacity and radiation is strong enough to trigger the RTI. For simplicity, we assume perfect thermal and dynamic coupling between gas and dust grains. The Planck $\kappa_\text{P}$ and Rosseland $\kappa_\text{R}$ mean opacities are given by
\begin{equation}
  \{\kappa_\text{P}, \kappa_\text{R}\} = \{0.1, 0.0316\}\, \left(\frac{T}{10\,\text{K}}\right)^2\,\text{cm}^2\,\text{g}^{-1} \, ,
\end{equation}
which approximately holds for dusty gas in LTE at $T \leq 150\,\text{K}$ \citep{Semenov2003}. At higher temperatures we simply limit both mean opacities to their values at 150\,K. We define a characteristic radiation temperature $T_\text{r} = (u / a_\text{B})^{1/4}$, which is nearly identical to the gas temperature $T$ except in thin layers of shock-heated gas, e.g. at the front of the wind. For simplicity, the gravitational acceleration $g$ is assumed to be constant (downward) and the radiation field is sourced by a constant flux $F_\ast$ at the lower boundary (upward). This flux defines a characteristic temperature $T_\ast = (F_\ast / a_\text{B} c)^{1/4}$, which leads to the definition of a corresponding characteristic sound speed $c_\ast = \sqrt{k_\text{B} T_\ast / \mu m_\text{H}}$, scale height $h_\ast = c_\ast^2 / g$, and sound crossing time $t_\ast = h_\ast / c_\ast$. As in previous studies we assume $\mu = 2.33$ and choose $T_\ast = 82\,\text{K}$, corresponding to $F_\ast = 2.54 \times 10^{13}\,\Lsun\,\text{kpc}^{-2}$ and $\kappa_{\text{R},\ast} = 2.13\,\text{cm}^2\,\text{g}^{-1}$. With these assumptions the levitation setup of radiation opposing gravity is characterized by two dimensionless numbers, the Eddington ratio
\begin{equation}
  f_{\text{E},\ast} = \frac{\kappa_{\text{R},\ast} F_\ast}{g c}
\end{equation}
and the optical depth
\begin{equation}
  \tau_\ast = \kappa_{\text{R},\ast} \Sigma_\ast \, ,
\end{equation}
where $\Sigma_\ast$ denotes the characteristic surface mass density. For the simulation presented in this paper we use $f_{\text{E},\ast} = 1/2$ and $\tau_\ast = 3$, which is sufficient to probe dynamically unstable coupling between radiation and gas.

\begin{figure*}
\centering
\includegraphics[width=.99\textwidth]{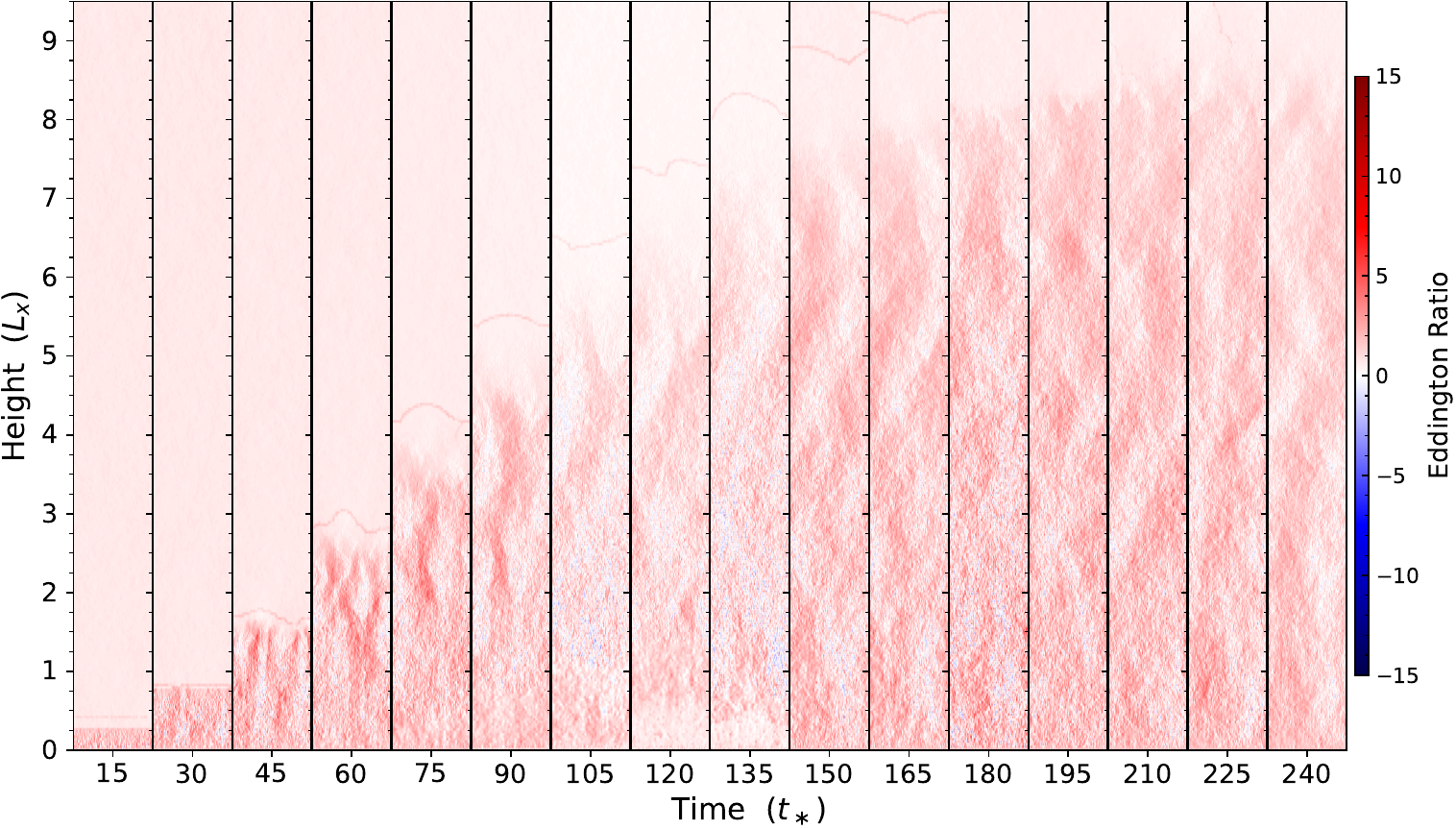}
\caption{Evolution of the vertical Eddington ratio $f_{\text{E},y}$, i.e. the ratio of radiative to gravitational accelerations. The volume is largely dominated by super-Eddington regions except for the dense filamentary clumps that descend until they are disrupted reminiscent of a recycling process.} \label{fig:levitation_f_Edd}
\end{figure*}

The initial setup corresponds to an isothermal atmosphere in hydrostatic equilibrium in the absence of radiation. Specifically, the temperature is $T = T_\ast$ and the vertical density profile is the exponential $\rho(y) = \rho_\ast \exp(-y/h_\ast)$, where $\rho_\ast = \Sigma_\ast / h_\ast$ is the characteristic density. We impose a density floor of $10^{-10}\,\rho_\ast$ and further induce a perturbation of the form
\begin{equation}
  \frac{\delta \rho}{\rho} = \frac{1 + \chi}{4} \sin\left(\frac{4\pi x}{L_x}\right) \, ,
\end{equation}
where $L_x = 512\,h_\ast$ is the simulation boxsize and $\chi$ is a random number uniformly distributed in $[-1/4, 1/4]$. Furthermore, no radiation is initially present. The boundary conditions are periodic in the $x$ direction, reflective at bottom ($y = 0$), and outflowing at the top ($L_y = 16 L_x$), tall enough that no gas is lost during the simulation. The initial mesh consists of a high-resolution $(0.5 h_\ast)^2$ Cartesian mesh at the bottom to resolve the high-density gas. The resolution is degraded slowly upwards until a minimum resolution of $(8 h_\ast)^2$ is reached. As the simulation progresses the mesh moves according to the local fluid flow, is regularlized where needed, and undergoes adaptive refinement and derefinement to approximately maintain cell volumes between $0.25$ and $64\,h_\ast^2$ with a target mass resolution of $\Sigma_\ast L_x / 1024^2$. Finally, we run with an adaptive convergence criteria of $\delta_\text{goal} = f_\text{goal} = 0.05$, a luminosity boosting exponent of $1/2$, and the traditional volume-based momentum scheme which is accurate because geometric sources are always at the boundary interfaces and the majority of cells are optically thin.

\subsection{Results}
Figures~\ref{fig:levitation_rho}, \ref{fig:levitation_T_rad}, and \ref{fig:levitation_f_Edd} respectively show the evolution of the normalized gas density $\rho / \rho_\ast$, radiation temperature $T_\text{r} / T_\ast$, and the vertical Eddington ratio $f_{\text{E},y}$, i.e. the ratio of radiative to gravitational accelerations for increasing times. The radiation field remains fairly smooth throughout the simulation and mirrors the density propagation as photons escape when they are able to reach the front of the wind. The trapping is high at early times ($t \approx 25\,t_\ast$), leading to efficient gas heating, which in turn increases the mean opacity to sustain an initial super-Eddington phase that quickly lifts the gas upwards. As Rayleigh--Taylor instabilities form, the vertical radiation coupling weakens and gravity slowly removes inertia ($t \approx 100\,t_\ast$). At later times, the photons regain sufficient trapping for levitation to continue ($t \approx 200\,t_\ast$) throughout the remainder of the simulation. Apparently, the radiation force continues to counterbalance gravity even as many of the dense filamentary structures stall out to what resembles a highly turbulent quasi-steady state.

\begin{figure}
\centering
\includegraphics[width=\columnwidth]{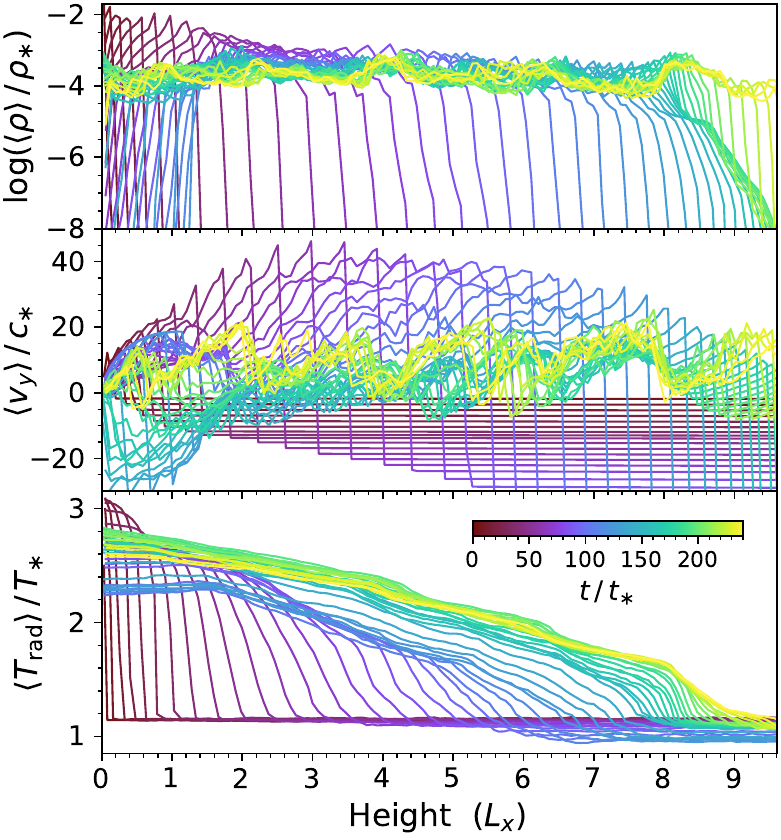}
\caption{Evolution of the normalized gas density $\rho / \rho_\ast$ (top), vertical mass-weighted velocity $v_y / c_\ast$ (middle), and radiation temperature $T_\text{r} / T_\ast$ (bottom) as one-dimensional profiles binned with a height resolution of $40\,h_\ast$ given in time intervals of $5\,t_\ast$. These quantities demonstrate the early propagation of the wind front and the highly turbulent quasi-steady-state at late times. The density becomes fairly uniform, the velocity develops wave-like behavior, and the temperature develops a characteristic gradient to support the continued levitation of the gas.} \label{fig:levitation_y}
\end{figure}

Figure~\ref{fig:levitation_y} provides complementary one-dimensional depictions of the evolving normalized gas density $\rho / \rho_\ast$, vertical (mass-weighted) gas velocity $v_y / c_\ast$, and radiation temperature $T_\text{r} / T_\ast$. These quantities are calculated in a conservative fashion by binning the gas particles with a height resolution of $40\,h_\ast$. The density shows the gas propagation and remains fairly uniform throughout time despite the significant elongation. The velocity is fastest near the front of the wind but develops a clear wave-like pattern with a peak-to-peak distance of approximately $2\,L_x$. Finally, the radiation temperature derived from the energy density allows us to better explore the cause of the apparent dimming in Figure~\ref{fig:levitation_T_rad}. These profiles clearly show that the radiation pressure is significantly relieved as newly emitted radiation escapes rather than continuing to build up ($t \approx 100\,t_\ast$). While the development of the RTI is critical for opening up low column density pathways, this is also a result of self-regulation as the initial driving phase was efficient enough to promote cooling via expansion of both the gas and radiation fields. As the wind stalls the radiation field is eventually reassembled to suppress substantial fallback. Interestingly, a characteristic temperature gradient develops to continue to support the gas at late times.

\begin{figure}
\centering
\includegraphics[width=\columnwidth]{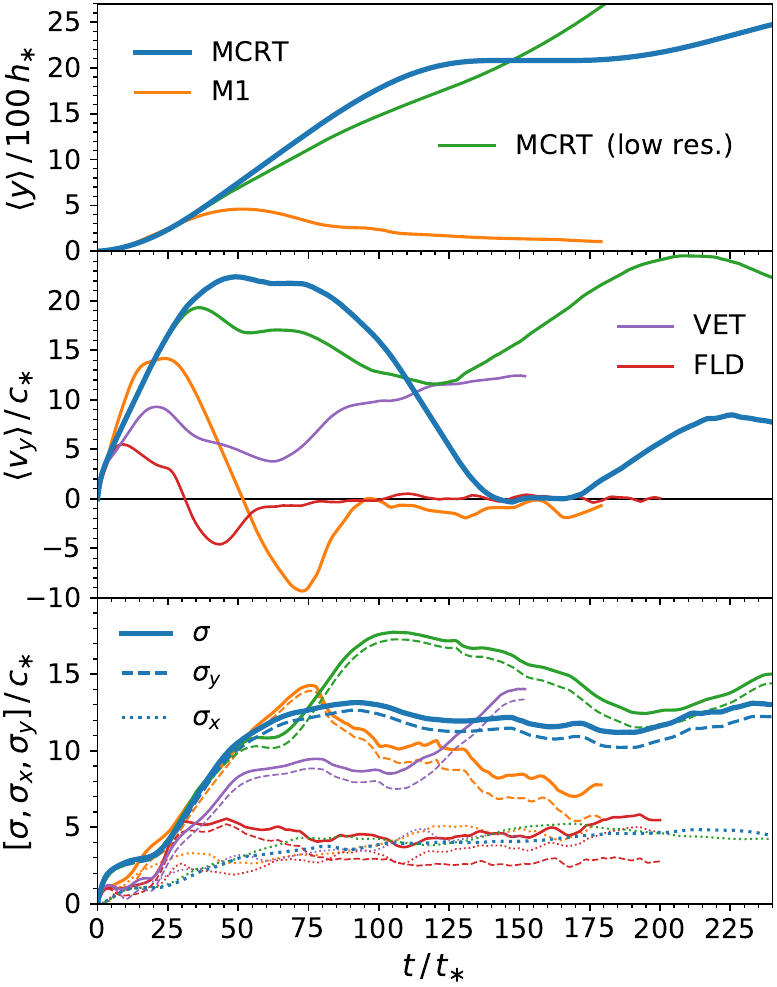}
\caption{Evolution of the global mass-weighted height $\langle y \rangle$ (top), vertical velocity $\langle v_y \rangle$ (middle), and velocity dispersion $\sigma$ (bottom). We compare our MCRT method (blue curves) to the M1 scheme of \ArepoRT\ \citep[orange curves;][]{Kannan2019} along with VET and FLD results \citep[purple and red curves;][]{Davis2014}, and find a significant difference in the ability to lift the gas upwards. The velocity reaches $\gtrsim 20\,c_\ast$ and despite the formation of instabilities remains positive even at late times. We also show results from a simulation with a target mass resolution that is four times lower than the main run (green curves).} \label{fig:levitation_time}
\end{figure}

\begin{figure}
\centering
\includegraphics[width=\columnwidth]{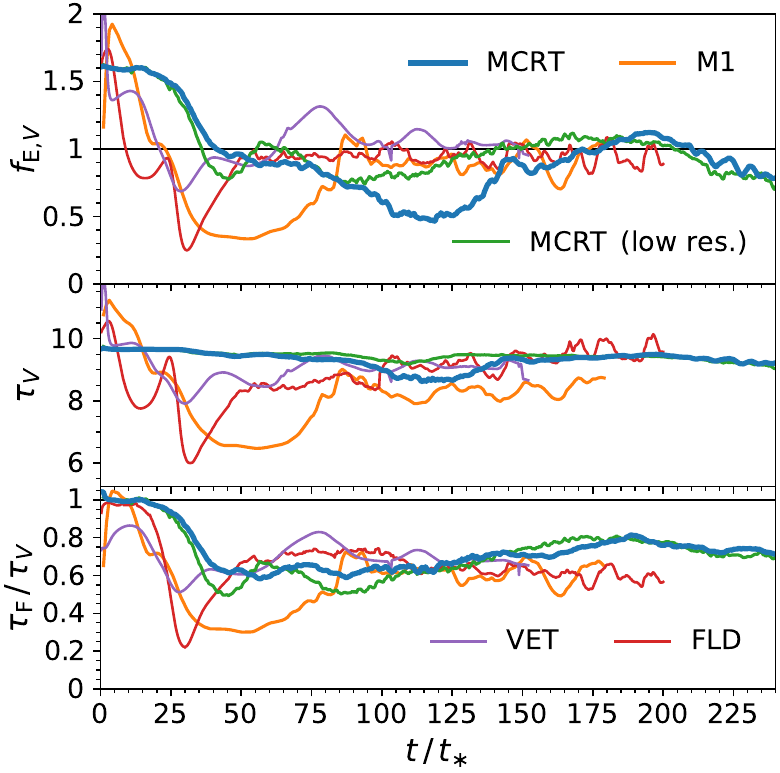}
\caption{Evolution of the global volume-weighted Eddington ratio $f_\text{E,V}$ (top), vertical optical depth $\tau_\text{V}$ (middle), and ratio of vertical flux- to volume-weighted optical depths $\tau_\text{F} / \tau_\text{V}$ (bottom). There is strong upward driving at early times ($t \lesssim 25\,t_\ast$), which is reduced but still relatively effective throughout the remainder of the simulation. We again compare our MCRT results (blue and green curves) to those obtained from the M1 (orange curves), VET (purple curves), and FLD (red curves) methods.} \label{fig:levitation_time_tau}
\end{figure}

Figure~\ref{fig:levitation_time} illustrates the evolution of the mass-weighted height $\langle y \rangle$ (top), vertical velocity $\langle v_y \rangle$ (middle), and velocity dispersion $\sigma$ (bottom) normalized to the appropriate characteristic scales. The initial gas acceleration resembles the result from \ArepoRT, while at late times follows the behavior of the previous VET and MCRT studies. We find a terminal velocity of $> 20\,c_\ast$, which is higher than all other results in the literature. The height reaches $\langle y \rangle \approx 2500\,h_\ast$ by the end of the simulation. The mass-weighted vertical velocity never falls significantly below zero, and remains highly turbulent for most of the run. To further explore the behavior of the radiation properties in Figure~\ref{fig:levitation_time_tau} we show additional globally averaged quantities. The net Eddington ratio (top) is defined as
\begin{equation}
  f_\text{E,V} = \frac{\sum f_{y,\text{rad},i}}{\sum g m_i} \, ,
\end{equation}
where the sums are over all gas cells $i$. During the initial acceleration phase the Eddington ratio is high $\gtrsim 1.5$ for a prolonged period but is mostly $\lesssim 1$ thereafter. The mean vertical optical depth (middle) is
\begin{equation}
  \tau_\text{V} = \frac{\sum k_{\text{R},i} V_i}{L_x} \, ,
\end{equation}
while the flux-weighted optical depth is
\begin{equation}
  \tau_\text{F} = \frac{\sum V_i \sum f_{y,\text{rad},i}}{L_x \sum f_{y,\text{rad},i}/k_{\text{R},i}} \, ,
\end{equation}
given in terms of the radiation force and absorption coefficient $k_\text{R}$. From Figure~\ref{fig:levitation_time_tau} it is clear that the mean trapping optical depth remains high throughout the simulation around $\tau_\text{V} \approx 9$. The bottom panel shows the ratio $\tau_\text{V} / \tau_\text{F}$, which is close to unity at the beginning but drops to lower values ($\approx 0.6$--$0.8$) for the remaining time as a reflection of the net traversal through lower opacity channels between filaments.

These results are quite interesting when compared to previous results. In particular, in the figures we highlight the comparison to the results from \ArepoRT, as this uses the same hydrodynamics code but with a second-order M1 closure scheme \citep[orange curves;][]{Kannan2019}. For additional comparison we also show results from the FLD (red curves) and VET (purple curves) methods \citep{Davis2014}. The mass-weighted gas height and velocity are significantly different, with the MCRT method launching a successful wind while M1 falls back down after an initial strong liftoff. This emphasizes the known result that moment-based methods and ray-tracing short/long characteristic RT methods differ in the long-term evolution. Specifically, the MCRT and VET methods avoid the characteristic long-term fate of the M1 and FLD methods to end up with turbulent gas that is gravitationally confined at the bottom of the domain. The gas dynamics can be significantly different between different radiative transfer methods, but the long-term results may also be sensitive to what happens during the initial acceleration phase as there are even differences between the previous MCRT implementation by \citet{Tsang2015}, although their results are similar to our lower resolution run. We also note that we followed their choice to cap the opacities $\kappa_\text{R,P}$ at their values at $T = 150\,\text{K}$, whereas all other authors allow the $\kappa \propto T^2$ scaling to arbitrary temperatures. Despite the opacity ceiling we still see the strongest overall RHD response compared to all other studies.

Finally, we briefly comment on a few numerical considerations. To determine how simulation resolution affects our results we performed an identical run maintaining cell volumes between $1$ and $64\,h_\ast^2$ with a target mass resolution of $\Sigma_\ast L_x / 512^2$, which is four times lower than the main run. The results are shown as green curves in Figures~\ref{fig:levitation_time} and \ref{fig:levitation_time_tau}, which exhibit identical behavior until about $35\,t_\ast$ when the instabilities are fully developed. Interestingly, at this point the chaotic behavior results in qualitatively different velocity histories with the lower resolution run being slightly slower during the first peak inertial phase but only falling to about $12\,c_\ast$ at $120\,t_\ast$ and rising with a second wind thereafter. This is similar to the MCRT results by \citet{Tsang2015}, which had resolution comparable to our low-resolution run ($1\,h_\ast^2$). \citet{Davis2014} also discuss the impact of resolution on the Eddington ratio and development of density inhomogeneities between their fiducial and lower ($1\,h_\ast^2$) resolution simulations. Neither of our studies demonstrate resolution convergence but we agree that higher resolution enhances the impact of radiative feedback in this test setup. More importantly, there is a larger difference between the treatment of radiation and hydrodynamics. To explore this further, we also ran a numerical experiment in which all photon packets are merged in each cell at the end of each time step. This emulates a crude moment-based approximation with poor flux preservation designed to test the importance of accurately representing intersecting rays from nonlocal radiation sources. In such simulations the wind fails with similar long-term behavior as the FLD and M1 results.

\section{Summary and Discussion}
\label{sec:summary}
In this paper, we have presented \ArepoMCRT, a novel implementation of a highly accurate Monte Carlo radiative transfer RHD method in the moving-mesh code \Arepo. The scheme uses a first-principles approach to sample the radiation field one photon trajectory at a time. The flow of a large but finite number of independently simulated photon packets provides a statistical representation of the collisionless radiation transport problem. The basic ideas are conceptually simple but the $\sqrt{N}$ rate of convergence requires variance reduction and importance sampling techniques to be competitive with other RHD methods in terms of accuracy to computational cost. We have incorporated many of the strategies employed throughout the community to overcome the inherent efficiency barriers inherent to MCRT. Beyond this, we have invested significant effort to develop concepts, discussion, and algorithms to meet the unique needs of \ArepoMCRT. In the long term, we intend to target a variety of multiple scattering problems relevant to astrophysical applications.

We tested our implementation on a variety of standard problems, and accurately reproduced the time-dependent transport and coupling effects in each instance. Specifically, we demonstrated the accurate diffusion of pulse and constant sources of radiation, simulating an infinite domain by tracking the tiling within periodic boxes. We also explore the $L^1$ error and hybrid DDMC speedup for these tests. We then derive a new analytic solution combining diffusion with homologous expansion to test our DDMC advection scheme. We resolved the evolution of rapidly cooling gas via thermal emission to a final state of radiative equilibrium. We also tested the hydrodynamical coupling by considering the formation of both sub- and super-critical radiative shocks.

Finally, we explored the ability of a trapped IR radiation field to accelerate a layer of gas in the presence of an opposing external gravitational field. We found persistent radiation-driven levitation even after the formation of Rayleigh--Taylor instabilities that create dense filaments and chimneys promoting the escape of photons. Our results are in agreement with previous VET and MCRT studies, as the photon directions are well preserved in these methods. On the other hand, the long-term behavior of the FLD and M1 closure moment-based methods is that of a highly turbulent quasi-steady state concentrated at the bottom of the domain. We note that due to efficient trapping at early times we obtain higher initial ballistic velocities than all previous studies. An important insight from our simulations is that the initial acceleration phase and the revitalized second wind are connected via self-regulation of the RHD coupling. As a consequence, the RHD implementation and simulation resolution are both crucially important when going beyond qualitative long-term effects.

We also emphasize that \ArepoMCRT\ can be used to post-process existing simulations to obtain accurate representations of radiation fields and emergent observables. In fact, MCRT is often the \textit{de facto} method of choice for multi-frequency three-dimensional radiative transfer calculations when considering scattered and reprocessed light. While we do not anticipate that \ArepoMCRT\ will replace existing post-processing pipelines, there is a natural place for built-in and on-the-fly tools with a common codebase tightly coupled to the original simulation. Specifically, the code inherits the efficient mesh construction, domain decomposition, and other strategies that are increasingly necessary for state-of-the-art hydrodynamics simulations, while at the same time introducing new algorithms and data structures that address the unique challenges of the computationally demanding nonlocal photon transport.

In the current state, it is not yet feasible to perform full galaxy formation simulations. However, the RHD solver will undergo continued development to address the performance needs of various applications. Importantly, this includes allowing compatibility with the individual time-stepping scheme of \Arepo\ and improving the strong scaling with the number of computational domains. Depending on the application, other strategies for MCRT physics algorithms and parallelization efficiency would also likely benefit our implementation \citep[e.g.][]{Harries2019,Michel-Dansac2020,VandenbrouckeCamps2020}. Still, even the high-resolution levitation setup can be run on a single compute node within a reasonable amount of time. This is partially due to the rapid adaptive convergence of the trapped radiation field and the ability to take longer time steps with the implicit Monte Carlo scheme, even when employing the full speed of light in transport calculations.

In future work, we plan to use this implementation to study timely problems in astrophysics. This includes the goal to study Lyman-$\alpha$ (Ly$\alpha$) radiation pressure with the first self-consistent three-dimensional Ly$\alpha$ RHD simulations. Previous studies have shown that resonant scattering by trapped Ly$\alpha$ photons can have a dynamical impact in dust-poor environments \citep{SmithRHD2017,Smith2019,Kimm2018}. We plan to study these phenomena with the Ly$\alpha$ radiative transfer functionality that is already implemented within \ArepoMCRT. The full RHD interface will also take advantage of the new resonant DDMC scheme proposed by \citet{SmithTsang2018} to break the efficiency barrier of frequency redistribution in this physical regime. Overall, it will be a valuable endeavor to push the limits of MCRT RHD schemes to provide an accurate and robust understanding of the role of radiation fields throughout the universe.

\acknowledgments

We thank the referee for insightful comments and suggestions which have improved the quality of this work. We thank David Barnes, Paul Duffel, Hui Li, Anna Rosen, Volker Springel, and Rainer Weinberger for fruitful discussions and advice related to this work. We thank Shane Davis for kindly sharing simulation data. Support for Program number HST-HF2-51421.001-A was provided by NASA through a grant from the Space Telescope Science Institute, which is operated by the Association of Universities for Research in Astronomy, Incorporated, under NASA contract NAS5-26555. M.V.\ acknowledges support through a NASA ATP grant NNX17AG29G, and NSF grants AST-1814053, AST-1814259, AST-1909831, and AST-2007355. This research was supported in part by the National Science Foundation under grant No. NSF PHY-1748958 and the Gordon and Betty Moore Foundation through Grant GBMF5076. Resources supporting this work were provided by the NASA High-End Computing (HEC) Program through the NASA Advanced Supercomputing (NAS) Division at Ames Research Center.

%% To help institutions obtain information on the effectiveness of their
%% telescopes the AAS Journals has created a group of keywords for telescope
%% facilities.
%
%% Following the acknowledgments section, use the following syntax and the
%% \facility{} or \facilities{} macros to list the keywords of facilities used
%% in the research for the paper.  Each keyword is check against the master
%% list during copy editing.  Individual instruments can be provided in
%% parentheses, after the keyword, but they are not verified.

% \vspace{5mm}
% \facilities{HST(STIS), Swift(XRT and UVOT), AAVSO, CTIO:1.3m,
% CTIO:1.5m,CXO}

%% Similar to \facility{}, there is the optional \software command to allow
%% authors a place to specify which programs were used during the creation of
%% the manuscript. Authors should list each code and include either a
%% citation or url to the code inside ()s when available.

% \software{astropy \citep{2013A&A...558A..33A},
%           Cloudy \citep{2013RMxAA..49..137F},
%           SExtractor \citep{1996A&AS..117..393B}
%           }

%% Appendix material should be preceded with a single \appendix command.
%% There should be a \section command for each appendix. Mark appendix
%% subsections with the same markup you use in the main body of the paper.

%% Each Appendix (indicated with \section) will be lettered A, B, C, etc.
%% The equation counter will reset when it encounters the \appendix
%% command and will number appendix equations (A1), (A2), etc. The
%% Figure and Table counter will not reset.

\newpage

\appendix

\section{Spherical coordinates}
\label{sec:spherical_coordinates}
Ray tracing in spherical coordinates can be reduced to finding the intersection between a line and sphere. We take $\bmath{r}$ and $\bmath{n}$ to be the photon position and direction before traversal such that the radius is bounded by inner and outer radii, i.e. $r \equiv \| \bmath{r} \| \in [r_{-},r_{+}]$. Thus, the length is determined by $\| \bmath{r} + \ell \bmath{n} \|^2 = r_{\pm}^2$, which admits solutions of
\begin{equation}
  \ell_{\pm} = -\mu r \pm \sqrt{r_{\pm}^2 - r_\text{min}^2} \, ,
\end{equation}
where the unnormalized radial cosine is $\mu r \equiv \bmath{n} \bmath{\cdot} \bmath{r}$, the impact parameter is $r_\text{min}^2 \equiv (1 - \mu^2) r^2$, and only positive lengths are intersections. In practice, if $\mu \geq 0$ or the inner discriminant $r_{-}^2 - r_\text{min}^2 \leq 0$ (but $\mu < 0$), then the photon packet traverses toward the outer shell boundary. Otherwise, we use the inner solution. To calculate a radial momentum similar to equation~(\ref{eq:momentum}) we integrate the outward contribution along the path, which yields
\begin{align} \label{eq:spherical_momentum}
  \Delta p_r &= \frac{k \varepsilon_k}{c} \int_0^\ell \hat{\bmath{r}} \bmath{\cdot} \left(\bmath{r} + \ell' \bmath{n}\right)\,\text{d}\ell' \notag \\
  &= \frac{k \varepsilon_k}{c} \int_0^\ell \frac{\ell' + \mu r}{\| \bmath{r} + \ell' \bmath{n} \|} \text{d}\ell' \notag \\
  &= \frac{k \varepsilon_k}{c} \left( \| \bmath{r} + \ell \bmath{n} \| - r \right) \equiv \frac{\tau_r \varepsilon_k}{c} \, ,
\end{align}
where $\tau_r \equiv k \Delta r$ is the net radial optical depth traversed. Unfortunately, there is no simple analytic formula when continuous absorption is taken into account by adding a factor of $e^{-\tau_\text{a}}$ inside the integral. However, in this case we can either (i) track the momentum flux through interfaces, (ii) switch to a probabilistic absorption scheme, or (iii) create an approximate numerical solution from
\begin{align}
  &\Delta p_r = \frac{k \varepsilon_k}{c} \int_{\mu r}^{\mu r + \ell} \frac{\ell' e^{k_\text{a} (\mu r - \ell')}}{\sqrt{\ell'^2 + r_\text{min}^2}} \text{d}\ell' \\
  &\approx \frac{k \varepsilon_k}{k_\text{a} c} e^{k_\text{a} (\mu r - r_\text{min})}
    \begin{bmatrix}
      \frac{e^{k_\text{a} (r_\text{min} - \ell')} (1 + k_\text{a} \ell') - (1 + k_\text{a} r_\text{min})}{k_\text{a} r_\text{min}} & \!\!\!|\ell'| < r_\text{min} \\
      \text{sgn}(\ell') \left(1 - e^{k_\text{a} (r_\text{min} - \ell')}\right) & \!\!\!|\ell'| \geq r_\text{min}
    \end{bmatrix}_{\mu r}^{\mu r + \ell} \notag
\end{align}
with the last line being an example of a simple but relatively accurate option. We emphasize that this is only consequential when there is significant absorption along the path segment. In fact, the center of energy distance along the ray is
\begin{equation}
  \langle \ell \rangle = \ell^{-1} \int_0^\ell \ell' e^{-k_\text{a} \ell'}\text{d}\ell' = \frac{\ell}{\tau_\text{a}^2} \left( 1 - (1 + \tau_\text{a}) e^{-\tau_\text{a}} \right) \, ,
\end{equation}
corresponding to a center of energy position of $\langle \bmath{r} \rangle = \bmath{r}_0 + \langle \ell \rangle \bmath{n}$. If the cell optical depth is greater than unity this can be highly skewed toward the origin of the ray as opposed to the optically thin midpoint of $\langle \ell \rangle \approx \ell/2 - \tau_\text{a} \ell/3 + \mathcal{O}(\tau_\text{a}^2)$. For completeness, we provide the first-order correction to equation~(\ref{eq:spherical_momentum}) as
\begin{align}
  \Delta p_r &\approx \frac{\tau_r \varepsilon_k}{c} + \frac{k_\text{a} k \varepsilon_k}{2 c} \bigg[ \mu r \Delta r - \ell (r + \Delta r) \notag \\
  &+ \frac{1}{2} r_\text{min}^2 \log\left( \frac{(1-\mu) (r + \Delta r + \ell + \mu r)}{(1+\mu) (r + \Delta r - \ell - \mu r)} \right) \bigg] \, .
\end{align}

\section{DDMC--MC boundary conditions}
\label{sec:DDMC_BCs}
We now demonstrate the validity of our new semi-deterministic DDMC momentum coupling scheme across extreme transitions. Specifically, this refers to equation~(\ref{eq:DDMC-momentum}) while also employing the hybrid IMC--DDMC boundary conditions described in Section~\ref{sec:hybrid-DDMC}. This is particularly important because opacity gradients often induce high radiative fluxes. Furthermore, transition regions are particularly sensitive to changes, so it is essential to accurately capture the local momentum coupling. We design a simple test to capture the relevant features of this numerical problem.

\begin{figure}
\centering
\includegraphics[width=\columnwidth]{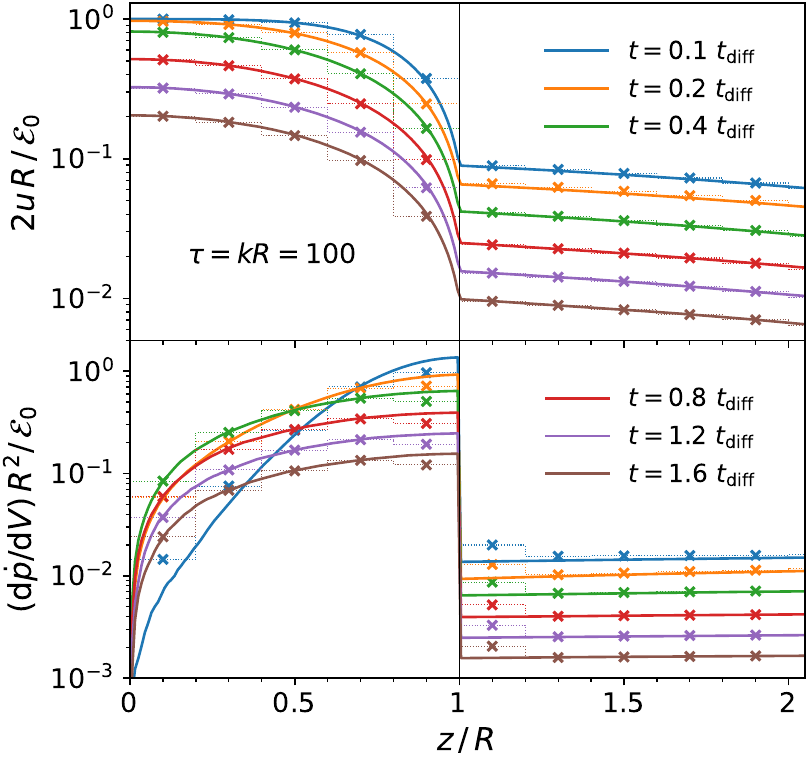}
\caption{Radiation energy density $u(z)$ and force density $\text{d}\dot{p}/\text{d}V$ for a tophat configuration with center-to-edge optical depth of $\tau = 100$ but a $100\times$ reduced density outside. This test demonstrates the validity of the semi-deterministic DDMC momentum coupling scheme across extreme DDMC--MC transitions. The curves illustrate the evolution at times of $t = \{0.1, 0.2, 0.4, 0.8, 1.2, 1.6\}\,t_\text{diff}$, where the diffusion time is $t_\text{diff} = k R^2 / 2 c$. The histograms employ the DDMC method a with resolution of $\Delta \tau = 20$, while the curves are the reference solutions with $\Delta \tau = 1$.} \label{fig:DDMC_BCs}
\end{figure}

The setup is that of a one-dimensional uniform slab with a center-to-edge optical depth of $\tau = k R = 100$. Outside the central tophat region the density drops by a factor of 100, representing an extreme transition layer before the photons escape freely at a radius of $3R$. We choose to restrict the photon propagation to pure forward--backward scattering, which promotes sufficiently rapid escape for a convenient cadence of distinct curves in our demonstration. We have verified that three-dimensional transport gives similar results. At $t = 0$ we initialize $10^7$ photon packets as a pulse source uniformly distributed throughout the central region, in an effort to enhance the flux near the transition layer. Fig.~\ref{fig:DDMC_BCs} shows the radiation energy and force density profiles over several representative times, $t = \{0.1, 0.2, 0.4, 0.8, 1.2, 1.6\}\,t_\text{diff}$, where in this case the diffusion time is $t_\text{diff} = k R^2 / 2 c$. The course-grained DDMC results with central cell resolutions of $\Delta \tau = 20$ are in excellent agreement with the MC reference solutions with central cell resolutions of $\Delta \tau = 1$. The slight discrepancy at early times is due to time averaging effects related to the lagged path-based estimators and the limitation of ignoring the flux time-derivative term in the diffusion closure relation (see equation~\ref{eq:simplified-RTE-1}). We partially mitigate this by employing a slightly shorter time step ($\Delta t = 0.05\,t_\text{diff}$) than the near-equilibrium times of interest ($t \gtrsim 0.1\,t_\text{diff}$). We conclude that DDMC transport in this regime is highly accurate and significantly more efficient than traditional MCRT.

\begin{figure}
\centering
\includegraphics[width=\columnwidth]{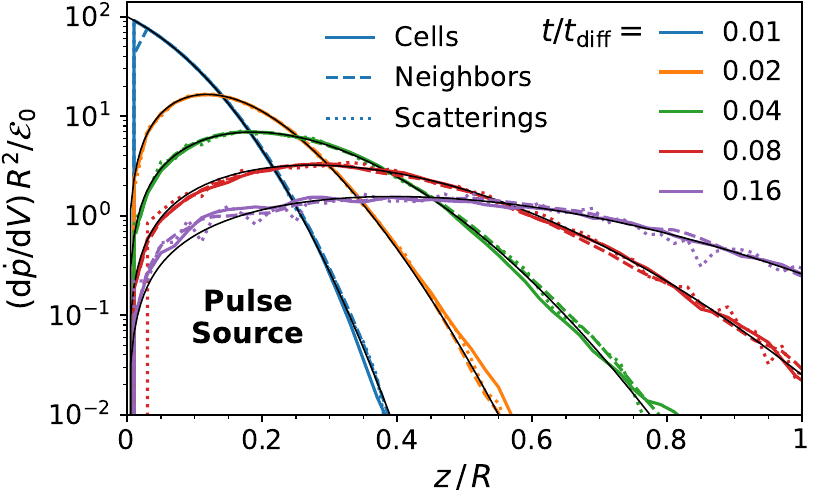}
\caption{Radiation force density $\text{d}\dot{p}/\text{d}V$ for a pulse source diffusing in a one-dimensional uniform medium over several doubling times, $t = \{1, 2, 4, 8, 16\} \times 10^{-2}\,t_\text{diff}$, where the diffusion time is $t_\text{diff} = k R^2 / 2 c$. Momentum schemes based on scattering events, path depositions, and neighbor integration all agree when the radiation field is well resolved. For direct comparison the analytic solution derived in Appendix~\ref{sec:neib_cons} is averaged over the lagged time step.} \label{fig:grey_diffusion_pulse_momentum}
\end{figure}

\section{Neighbor Momentum Conservation}
\label{sec:neib_cons}
We now validate the efficacy of the neighbor momentum coupling scheme to conserve momentum compared to the cell-integrated method. We emphasize that problems arise only when individual cells are optically thick but that such scenarios are of common occurrence even in hydrodynamics simulations with state-of-the-art resolution. We refer to \citep{Hopkins2019} for a detailed discussion regarding non-scattering radiation pressure and to Section~\ref{sec:momentum} above for our proposed general MCRT solution. The setup we explore is the one-dimensional slab version of the test from Section~\ref{sec:grey_diffusion}. The simulated domain is large enough so essentially no photons escape and although the system is scale free for concreteness we choose a characteristic optical depth of $\tau = k R = 500$. We choose to restrict the photon propagation to pure forward--backward scattering, which avoids transverse geometric vector cancellation and simplifies the derivations for reference solutions. Specifically, the evolution of the radiation energy density is governed by a diffusion equation $\partial u / \partial t = (c/k) \partial^2 u / \partial z^2$ and the solution given an initial point source impulse of energy $\mathcal{E}_0$ is $\tilde{u} = e^{-\tilde{z}^2/2\tilde{t}} / (2 \pi \tilde{t})^{1/2}$, where we have rescaled into dimensionless units with radius $\tilde{r} = r / R$, time $\tilde{t} = t / t_\text{diff}$ with diffusion time $t_\text{diff} = k R^2 / 2 c$, and energy density $\tilde{u} = u\,R/\mathcal{E}_0$. Therefore, the resulting force density is $\text{d}\dot{p} / \text{d}V = k \bmath{F} / c = -\bmath{\nabla} u = (\mathcal{E}_0 / R^2) \tilde{z} \tilde{u} / \tilde{t}$. The total instantaneous force integrated over all space is $\dot{p}_\text{tot} = 2 \int_0^\infty (\text{d}\dot{p} / \text{d}z)\,\text{d}z = (\mathcal{E}_0 / R) (2 / \pi \tilde{t})^{1/2}$, such that the cumulative momentum up to a given point in time is $p_\text{tot} = \int_0^t \dot{p}_\text{tot}\,\text{d}t' = (\tau \mathcal{E}_0 / c) (2 \tilde{t} / \pi)^{1/2}$. We note that similar expressions may also be derived for spherical geometry.

\begin{figure}
\centering
\includegraphics[width=\columnwidth]{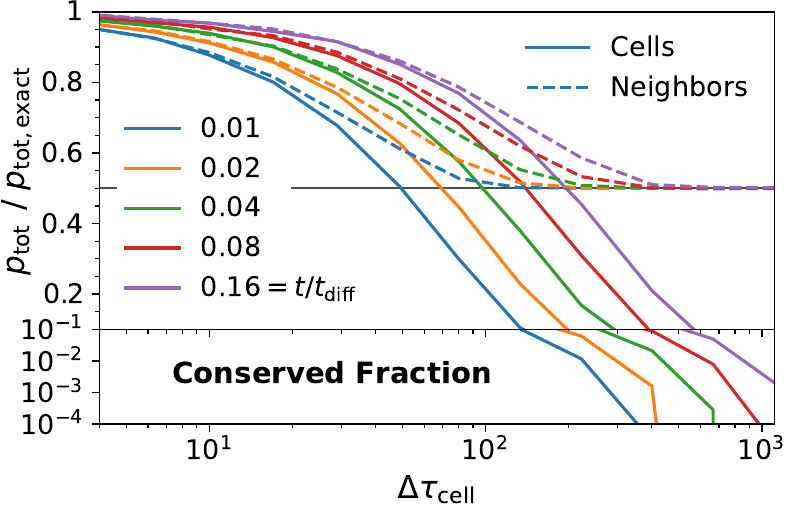}
\caption{Momentum conservation as a function of cell optical depth resolution based on integrating the radiation force density over space and cumulative time and dividing by the exact analytic value of $p_\text{tot} = (\tau \mathcal{E}_0 / c) (2 \tilde{t} / \pi)^{1/2}$. While standard volume averaged schemes can result in significant losses the neighbor method is guaranteed to conserve momentum within a factor of two.} \label{fig:grey_diffusion_pulse_momentum_issue}
\end{figure}

\begin{figure}
\centering
\includegraphics[width=\columnwidth]{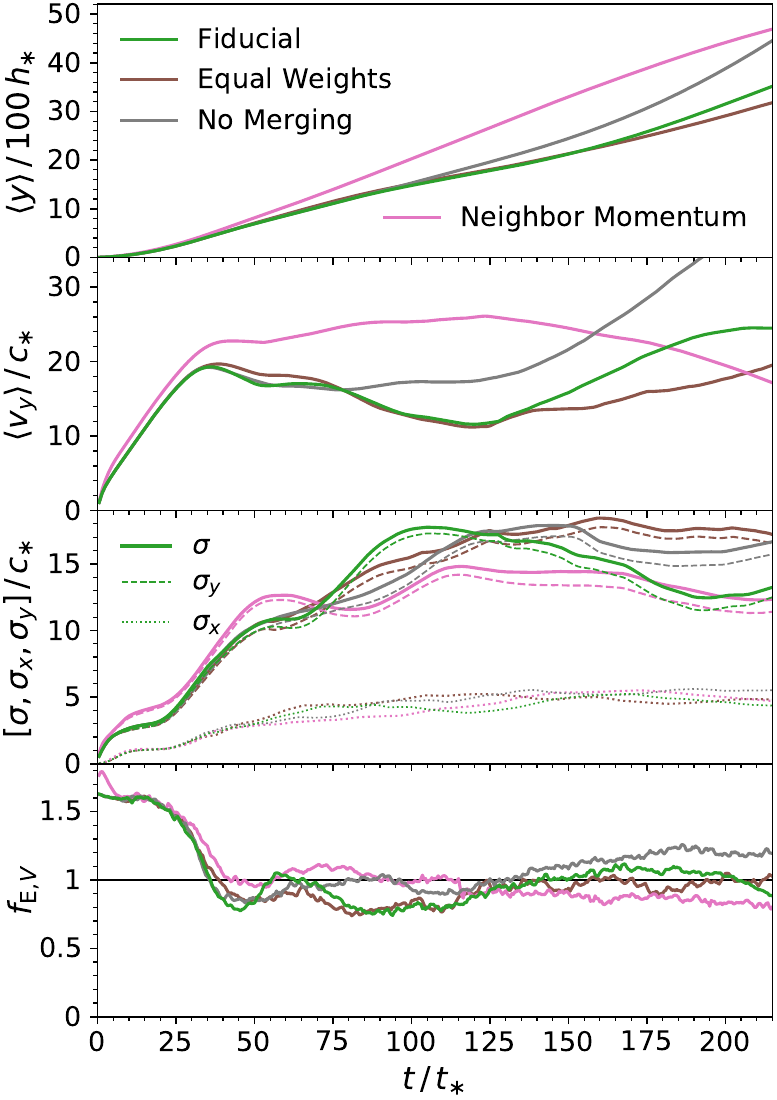}
\caption{Evolution of the global mass-weighted height $\langle y \rangle$ (top), vertical velocity $\langle v_y \rangle$ (second), velocity dispersion $\sigma$ (third), and volume-weighted Eddington ratio $f_\text{E,V}$ (bottom). We compare the fiducial MCRT model with luminosity boosting, merging, and splitting as employed in Section~\ref{sec:levitation} above (green curves) to identical simulations but with equal photon weights (brown curves), no merging or splitting (gray curves), and employing the neighbor momentum scheme (pink curves), all at lower resolution ($1\,h_\ast^2$). Although they are all within the expected agreement, this demonstrates that implementation details can also be important.} \label{fig:levitation_time_variations}
\end{figure}

We demonstrate the success of the three momentum implementations illustrated in Fig.~\ref{fig:momentum_diagram}: (i) microphysical scattering based exchange as described in \citet{Tsang2015}, (ii) volume integration of path depositions, and (iii) neighbor-based coupling accounting for physical self-cancellation and outwardly oriented propagation. Fig.~\ref{fig:grey_diffusion_pulse_momentum} shows the radiation force density profile over several doubling times, $t = \{1,2,4,8,16\} \times 10^{-2}\,t_\text{diff}$, employing $10^5$ photon packets so statistical variations are apparent. The simulation provides excellent agreement with the exact analytical solution, although we note that path-based estimators represent averages over discrete time steps and cell volumes so we plot the analytic solution including the appropriate time lag and bin integration. Finally, in Fig.~\ref{fig:grey_diffusion_pulse_momentum_issue} we show the ratio of simulated to exact cumulative momenta, which demonstrates the failure of cell-integrated methods (of all varieties) to conserve momentum within optically thick cells. We place the point source at a cell center within the uniform grid and run simulations with varying resolutions. The main losses occur in the source cell at early times but persist to a noticeable degree until the diffusion is well resolved, i.e. $t \gg k \Delta z^2 / c$. In contrast, the neighbor momentum scheme conserves at least half of the momentum even when the radiation field is confined within a single cell. We note that the factor of two arises from systematic over cancellation, which can potentially be avoided by increasing the momentum imparted analogous to a closure relation. However, we prefer not to introduce this additional factor, which could erroneously apply too much momentum in other settings. By splitting the momentum between host and neighbor cells we successfully overcome the order of magnitude losses inherent to standard MCRT momentum coupling at poor optical depth resolutions.

\section{Levitation control experiments}
\label{sec:control}
In this section we briefly explore the impact of various algorithm choices in the context of the levitation problem. We re-run the low-resolution setup under the following scenarios: (i) the luminosity exponent is 1, which samples equal weight photons in an unbiased fashion, (ii) with merging and splitting turned off, and (iii) with the neighbor-based momentum scheme activated. We note that these are the only changes to the simulation setup. The results indicate that luminosity boosting has a minor impact on the late time behavior after the development of large density fluctuations in the turbulent gas. On the other hand, the effect of merging and splitting is more noticeable as these optimizations introduce artificial momentum and energy losses in the radiation field. Finally, the neighbor-based momentum method results in a significant boost in upward lift and gas expulsion, similar to the higher resolution fiducial run. We interpret this as an indication that the RHD physics is not converged at these low resolutions. We summarize these findings in Figure~\ref{fig:levitation_time_variations}.

%% For this sample we use BibTeX plus aasjournals.bst to generate the
%% the bibliography. The sample63.bib file was populated from ADS. To
%% get the citations to show in the compiled file do the following:
%%
%% pdflatex sample63.tex
%% bibtext sample63
%% pdflatex sample63.tex
%% pdflatex sample63.tex

\bibliography{biblio}{}
\bibliographystyle{aasjournal}

%% This command is needed to show the entire author+affiliation list when
%% the collaboration and author truncation commands are used.  It has to
%% go at the end of the manuscript.
%\allauthors

%% Include this line if you are using the \added, \replaced, \deleted
%% commands to see a summary list of all changes at the end of the article.
%\listofchanges

\end{document}